\documentclass[manuscript,screen]{acmart}
\usepackage{hyperref}
\usepackage{hyperxmp}
\usepackage{enumitem}
\usepackage{subfigure}
\usepackage{pifont}
\usepackage{subcaption}
\usepackage{graphicx}
\usepackage{ulem}
\usepackage{multirow}
\usepackage{subfigure}
\usepackage{amsmath}
\usepackage{cases}
\usepackage{colortbl}
\usepackage{tcolorbox}
\tcbuselibrary{breakable}
\usepackage{listings}
\tcbuselibrary{listings}
\usepackage[font=normalsize,skip=5pt]{caption}
\usepackage[ruled,linesnumbered,vlined]{algorithm2e}
\usepackage{listings}
\usepackage{ulem}
\usepackage{color}
\definecolor{pblue}{rgb}{0.13,0.13,1}
\definecolor{pgreen}{rgb}{0,0.5,0}
\definecolor{pred}{rgb}{0.9,0,0}
\definecolor{pgrey}{rgb}{0.46,0.45,0.48}
\usepackage{xspace}
\usepackage{tikz}
\usepackage{pgfplots}
\usepackage{tcolorbox}
\usepackage{colortbl}
\pgfplotsset{compat=1.3}
\makeatletter
\DeclareRobustCommand\onedot{\futurelet\@let@token\@onedot}
\def\@onedot{\ifx\@let@token.\else.\null\fi\xspace}

\makeatother
\AtBeginDocument{%
  \providecommand\BibTeX{{%
    \normalfont B\kern-0.5em{\scshape i\kern-0.25em b}\kern-0.8em\TeX}}}

\begin{document}

\title{Towards Exception Safety Code Generation with Intermediate Representation Agents Framework}

% \author{Jia Li}

% \affiliation{%

\author{Xuanming Zhang}
\affiliation{
  \institution{University of Wisconsin-Madison}
  \city{Madison}
  \country{United States}
}
\affiliation{
  \institution{Stanford University}
  \city{California}
  \country{United States}
}
\email{xzhang2846@wisc.edu}
\author{Yuxuan Chen}
\affiliation{
  \institution{Tsinghua University}
  \city{Beijing}
  \country{China}
}
\email{chenyuxu21@mails.tsinghua.edu.cn}
\author{Yuan Yuan}
\affiliation{
  \institution{Beihang University}
  \city{Beijing}
  \country{China}
}
\email{yuan21@buaa.edu.cn}
\author{Minlie Huang}
\affiliation{
  \institution{Tsinghua University}
  \city{Beijing}
  \country{China}
}
\email{aihuang@tsinghua.edu.cn}

\renewcommand{\shortauthors}{Zhang et al.}

\renewcommand{\paragraph}[1]{\vskip 0.03in \noindent {\bf #1.}}
\newcommand{\zhz}[1]{\textcolor{red}{zhz: #1}}
\newcommand{\lz}[1]{\textcolor{blue}{#1}}
\newcommand{\citeay}[1]{\citeauthor{#1}~\cite{#1}}
\def\poisonername{{\sc CodePoisoner}\xspace}
\def\detectorname{{\sc CodeDetector}\xspace}

\begin{abstract}
Large Language Models (LLMs) often struggle with robust exception handling in generated code, leading to fragile programs that are prone to runtime errors. We propose Seeker, a novel multi-agent framework that enforces exception safety in LLM-generated code through an Intermediate Representation (IR) approach. Seeker decomposes exception handling into five specialized agents – Scanner, Detector, Predator, Ranker, and Handler – that collaboratively analyze code, detect fragile segments, retrieve best-practice exception strategies, and inject robust handling code. We also introduce Common Exception Enumeration (CEE), a comprehensive knowledge base derived from official documentation, technical practices, and real-world code, to standardize exception handling strategies. Seeker also incorporates a Deep Retrieval-Augmented Generation (Deep-RAG) algorithm to efficiently navigate the exception inheritance hierarchy, cutting down search overhead by ~93\% while improving accuracy in identifying relevant exceptions. We evaluate Seeker on 15 open-source Java projects and multiple benchmarks. Seeker outperforms state-of-the-art baselines, improving exception handling precision by up to 37\% and overall code robustness by ~38\% as measured by expert code review. It significantly closes the gap between LLM and human developers in exception management, achieving a 28\% success rate on real-world issue fixes (SWE-bench) versus 19\% by prior methods. Our framework preserves functional correctness of code while proactively handling errors, demonstrating a practical, generalizable solution for safer code generation. In this paper, we discuss the novelty of using intermediate representation and multi-agent collaboration for exception handling, and outline how Seeker can be extended to other programming languages and complex software engineering tasks, aligning LLM-generated code with industrial standard.
\end{abstract}


\maketitle

\section{Introduction}
\label{sec1}
Recent advances in AI-driven code generation have enabled LLMs to produce substantial portions of software code \cite{deepseekcoder,codellama,starcoder}. However, ensuring \textbf{exception safety} – the ability of code to handle runtime errors gracefully – remains a critical challenge \cite{deveval}. LLM-generated code often contains \textbf{fragile exception handling}: it may ignore important failure cases, catch exceptions too generically, or apply misguided error-handling strategies. These shortcomings degrade code robustness and can introduce security vulnerabilities \cite{swebench, securityeval, LLM1, gptj}. Prior studies have observed developers frequently overusing generic exceptions and providing inadequate handling blocks, issues that not only reduce system stability but also taint the training data for code models. Addressing exception handling in code generation is thus essential for both immediate software reliability and the long-term improvement of coding assistants.

Human developers generally outperform LLMs in exception management by leveraging domain knowledge and structured reasoning. For instance, expert developers prioritize specific exceptions (e.g., using \texttt{SSLHandshakeException} rather than a generic \texttt{IOException} in security modules) and dynamically adjust handling granularity based on context. They also systematically navigate exception inheritance hierarchies – for example, understanding Java’s ~433 exception classes organized in branches – to catch more precise exception types \cite{should-fine}. By contrast, out-of-the-box LLMs (e.g., GPT-4) often handle exceptions insensitively or inaccurately, leading to a reported 63\% performance gap between GPT-4 and senior developers in exception-related tasks. This gap highlights the need for approaches that can inject expert-like reasoning into LLM-based code generation.

\begin{figure}[t]
  \centering
  \includegraphics[width=0.69\linewidth]{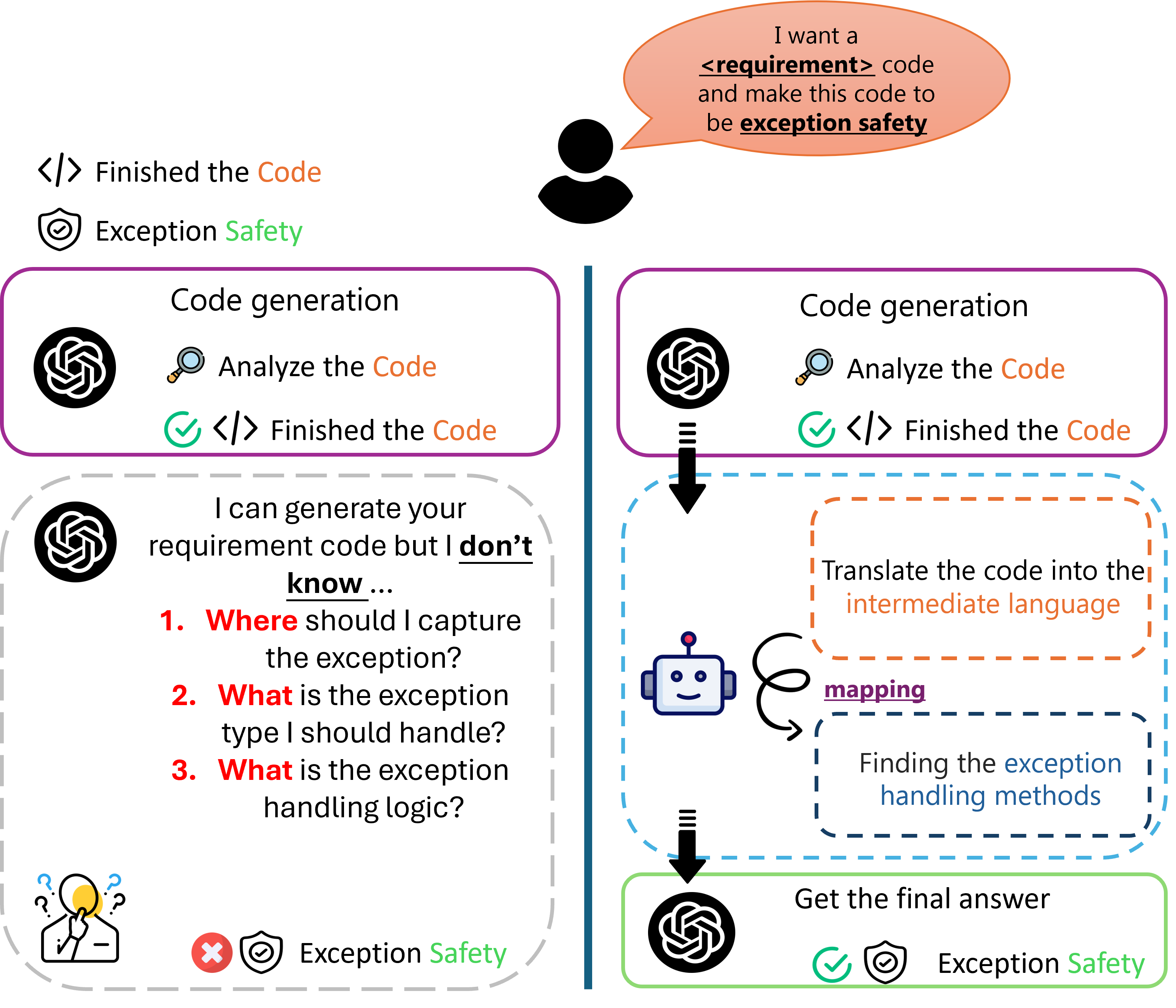}
  \caption{\textbf{ Overview of the Intermediate Representation (IR) agents (Right) Compared with Traditional Code Generation Approaches (Left) in Exception-Safe Code Generation Tasks.}  Seeker framework leverages IR agents to perform dynamic analysis, transformation, and optimization of code to ensure robust exception handling. In contrast, traditional approaches often rely on static error-handling routines and lack comprehensive analysis for exception safety.}
  \label{fg1}
  \vskip -.2in
\end{figure}

\begin{figure*}
\centering
    \subfigure[Preliminary following the four settings and strategies of expert reasoning. The vertical axis represents the evaluation score of human code review. \label{fig1.1}]{\includegraphics[width=0.45\textwidth]{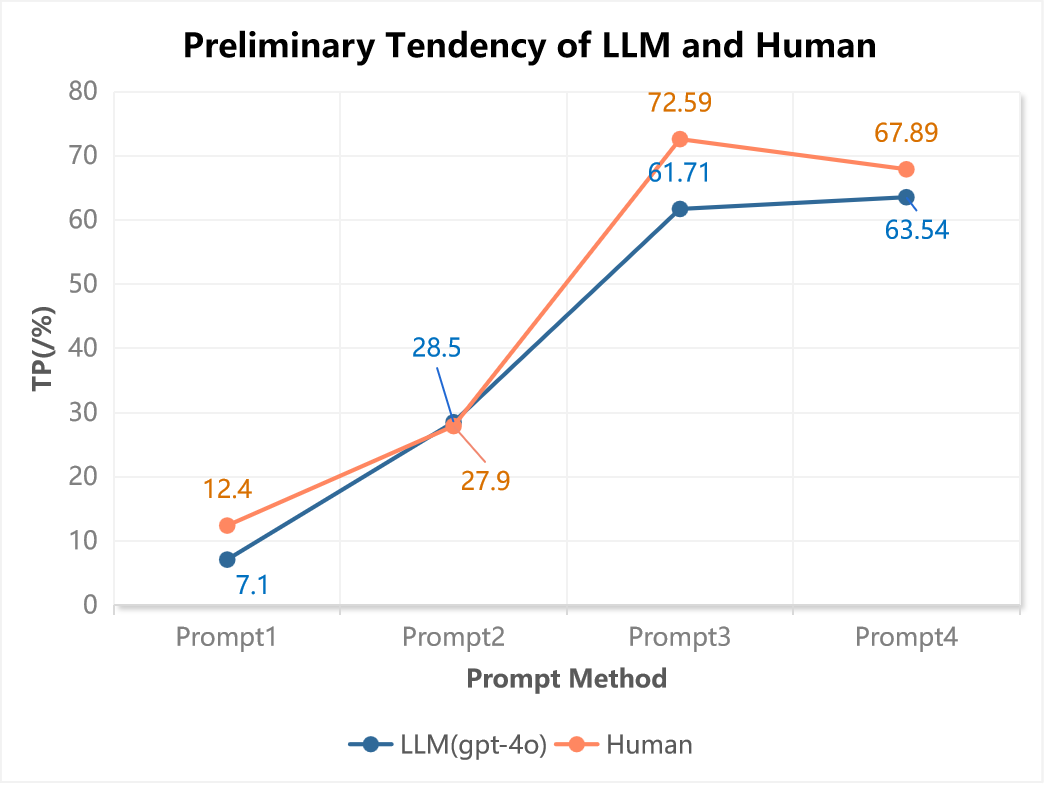}}
    \hfill
    \subfigure[A schematic diagram of human developers who well-performed in exception handling.\label{fig1.2}]{\includegraphics[width=0.45\textwidth]{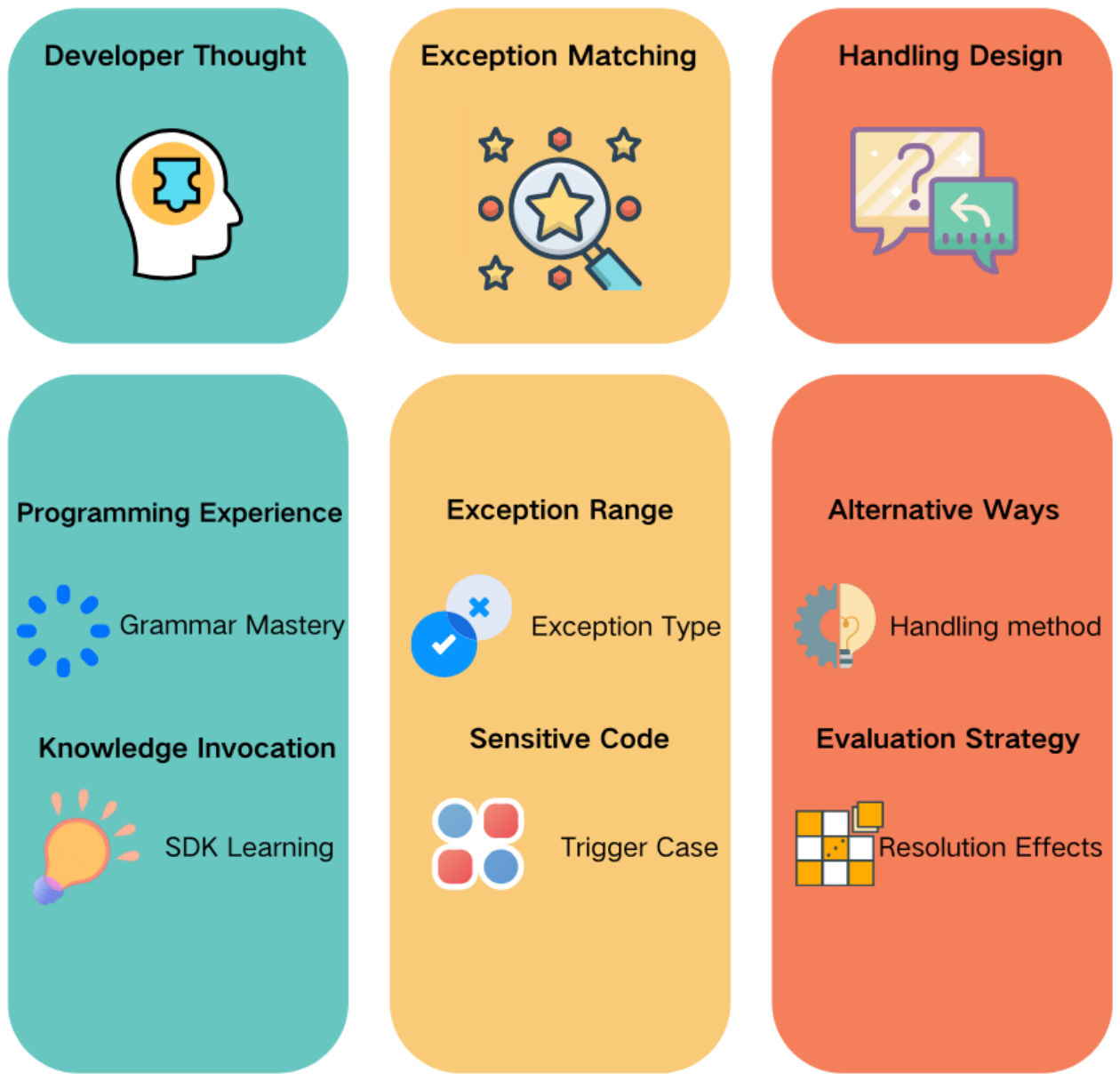}}
    % \subfigure[]\label{fig1.1}{\includegraphics[width=0.45\textwidth]{figure/preliminary.png}}
    % \hfill
    % \subfigure[]\label{fig1.2}{\includegraphics[width=0.45\textwidth]{figure/CoT.pdf}}
\caption{(a) Comparison of LLM and human exception handling performance as prompts evolve from \ding{182} Coarse-grained Reminding, to \ding{183} Fine-grained Reminding, \ding{184} Fine-grained Inspiring, and \ding{185} Fine-grained Guiding. Results show a clear mitigation effect, where increasingly detailed and context-rich prompts significantly improve handling quality. (b) How expert human developers integrate programming expertise, domain knowledge, fine-grained exception hierarchies, and adaptive strategies to achieve robust exception management.}
\label{fig1}
\vskip -0.1in
\end{figure*}

In this work, we redefine the research question to: \textit{How can we systematically eliminate developer-like defects in exception handling – including insensitive detection of fragile code, inaccurate exception capture, and distorted handling logic – in LLM-generated code?} To answer this, we propose using an \textbf{Intermediate Representation (IR) paradigm} that explicitly codifies expert exception-handling reasoning for the LLM. In preliminary study, we identified four levels of prompting that increasingly guide the LLM toward better exception handling: (1) \textit{Coarse-grained reminding} (basic alerts like “handle possible errors here”), (2) \textit{Fine-grained reminding} (type-specific suggestions, e.g., “catch \texttt{SQLException} instead of a generic \texttt{Exception}”), (3) \textit{Fine-grained inspiring} (contextual scenarios and risk analysis in prompts), and (4) \textit{Fine-grained guiding} (structured, step-by-step resolution strategies). Our experiments showed that providing more structured and detailed guidance (moving from coarse to fine-grained guiding prompts) significantly improves LLM exception handling performance, as shown in Figure \ref{fig1.1}, narrowing the gap to human expertise. This motivated the design of a specialized intermediate representation for exception handling knowledge, enabling the LLM to reason in a step-by-step, \textbf{human-like manner} \cite{metamind,seeker}.

\textbf{We present Seeker, a multi-agent collaboration framework that embodies this IR approach to exception safety code generation.} Seeker orchestrates five agents, each handling a distinct aspect of the exception-handling process, and integrates an external knowledge base of technical practices. The key contributions of our work are summarized as follows:

\begin{itemize} 
    \item \textbf{Seeker Multi-Agent Framework:} We design a novel agent-based architecture that breaks down exception handling into five stages – \textbf{Scanner, Detector, Predator, Ranker, and Handler} – each performed by a specialized agent. This modular design enables interpretable, stepwise code analysis and transformation, prioritizing proactive robustness. The agents collaborate to detect fragile code and inject appropriate error-handling constructs, effectively encoding a chain-of-thought for exception management within the LLM. Our framework is \textbf{language-agnostic} and easily extensible due to the separation of concerns among agents.
    \item \textbf{Common Exception Enumeration (CEE):} We develop a comprehensive knowledge base of exception handling strategies, called CEE, derived from trusted sources. CEE organizes exceptions into a hierarchy (based on language-specific inheritance, e.g., Java’s Exception classes) and for each exception type provides structured information: typical scenarios when it arises, important properties (e.g. checked vs unchecked), and recommended handling logic. By standardizing technical practices for hundreds of exceptions, CEE serves as an \textbf{explainable IR} for exceptions, guiding the Predator and Ranker agents in identifying and selecting proper handling strategies. CEE is built by integrating authoritative documentation (e.g., JDK specs), enterprise-level practice guides, and mining of real-world code repositories. This knowledge base not only boosts our system’s performance but is also a resource for developers, promoting community knowledge-sharing on exception handling\footnote{https://common-exception-enumeration.github.io/CEE/}.
    \item \textbf{Deep Retrieval-Augmented Generation (Deep-RAG):} We introduce Deep-RAG, a retrieval-augmented generation algorithm that efficiently navigates the complex exception hierarchy using contextual cues. Deep-RAG assigns scenario-based labels to branches of the exception tree and uses few-shot learning to generalize these labels. Given a code context (e.g., a code snippet and a detected fragile operation), Deep-RAG quickly focuses on the relevant branch of the exception hierarchy and retrieves the most pertinent exception types and handling templates from CEE. This significantly reduces the search space and overhead (by ~93\% compared to naive retrieval over all 433 Java exception types) while improving accuracy in pinpointing the correct exceptions to handle. Deep-RAG thus optimizes the knowledge retrieval process for the Predator/Ranker agents, enabling real-time use of the CEE knowledge in code generation. Notably, our Deep-RAG method supports multi-pattern exception handling (such as multi-catch blocks for different exceptions) by identifying groups of exceptions that can be handled together, and it can adapt to new domains by re-labeling the hierarchy.
    \item \textbf{Empirical Evaluation and Impact:} We conduct extensive experiments to validate Seeker. On a benchmark of 15 open-source Java projects (spanning 2019–2024) with 750 identified fragile code segments, Seeker consistently outperforms several baselines – including naive prompting, web search augmentation, and recent advanced methods like KPC \cite{kpc} and FuzzyCatch \cite{baseline3} – across multiple metrics. Seeker achieves \textbf{91\% coverage} of exception-prone code (vs. ~56\% by the best baseline) and \textbf{79\% accuracy} in catching the correct exception types (vs. ~43\% baseline), while its generated code closely matches expert-written fixes (Edit Similarity 0.64) and earns high approval in automated code reviews (92\% Code Review Score). These improvements translate to a \textbf{37\% precision gain} in detecting and handling exceptions and a \textbf{38\% boost in overall code robustness} over prior state-of-the-art. We also perform ablation studies to quantify the contribution of each agent in the framework, model variation tests using different underlying LLMs, and knowledge base analyses comparing performance with vs. without CEE. Finally, we demonstrate Seeker’s generality via additional benchmarks: on SWE-bench \cite{swebench} (real bug fixes from GitHub issues) Seeker solves ~28\% of issues (versus 19\% by previous agent-based approach), and on CoderEval \cite{codereval} (pragmatic code generation tasks) Seeker improves Codex’s success rate from 27.8\% to 38.2\%. These results highlight Seeker’s practicality and its potential to augment any code generation model with safer exception handling.
    \item \textbf{Alignment with Technical Practices:} Through multiple design choices, we explain why Seeker focuses on Java (a language with one of the most complex exception hierarchies and strict handling requirements) as an initial target, and why we adopt the try-catch mechanism as the primary handling technique. Java’s 433+ exception classes and pervasive use of checked exceptions make it an ideal proving ground – success here indicates our approach can scale to other languages with simpler or differently structured exception systems. We also analyze how Seeker’s IR-guided prompting leads developers (and LLMs) toward “good practice” exception handling patterns, in contrast to the ad-hoc or overly broad handling often seen in practice: Seeker’s role not only as a code generator but also as a teaching tool that infuses industry-standard robustness into development workflows.
\end{itemize}

\section{Motivating Examples}
\label{sec:motivating_examples}

Before detailing our approach, we present two sample investigations that shaped the motivation of Seeker, as shown in Figure \ref{fig2.1} and \ref{fig2.2}. We also quantify the performance gap between LLMs and human developers in exception handling tasks to motivate our focus.

We conducted an empirical study to measure how well current LLMs handle exceptions compared to experienced human developers. We curated a dataset of 12,893 code samples from 25 well-maintained enterprise repositories (spanning Java and Python projects such as Apache Commons, Spring Framework, etc.). Within these, we identified 2,147 fragile code segments – portions of code that either lacked necessary exception handling or contained error-prone handling (e.g., overly broad catches, empty catch blocks). Identification was done via a hybrid manual and automated review, including tracing historical commits to find where exceptions caused bugs or were fixed.

Analyzing these cases, we found two prevalent categories of developer mistakes in exception handling:

\ding{182} \textbf{Inaccurate Capture ($\approx$38.6\%)} - catch blocks that caught exceptions at the wrong level of specificity or caught the wrong exception type altogether. For example, catching a generic \texttt{Exception} or \texttt{Throwable} when a more specific exception was appropriate, or catching an exception that would never be thrown in that context.

\ding{183} \textbf{Distorted Handling ($\approx$41.2\%)} – handling logic that was ineffective or harmful. This includes empty catch blocks (swallowing exceptions), logging without addressing critical failures, or providing fallback behaviors that introduced new bugs.

These issues point to a lack of standardized approach among developers for dealing with exceptions, especially for uncommon error cases. We then compared how an advanced LLM (GPT-4, prompted in a straightforward manner) fared on these fragile code segments versus human fixes. The LLM was asked to generate exception-safe versions of the code. The results revealed a 63\% performance gap in robustness between the LLM’s output and the human-corrected code. Specifically, the LLM often failed in the same areas as junior human developers:

\ding{182} \textbf{Missed context-specific domain knowledge.} For instance, it didn’t prioritize handling \texttt{SSLHandshakeException} in a security-critical module, treating it like any other \texttt{IOException}, whereas human experts knew this was a critical exception needing its own handling.

\ding{183} \textbf{Lacked adaptive strategies.} Experts adjusted how they handled exceptions depending on the call stack depth or the component – e.g., a deep utility function might catch and rethrow (to let higher levels handle), while a top-level API endpoint would catch and respond to the user. The LLM did not dynamically adapt to such context as effectively.

\ding{184} \textbf{Struggled with hierarchical reasoning.} In cases where multiple exception types share inheritance (like I/O exceptions vs more specific socket exceptions), the LLM either handled only a generic parent or tried to handle everything individually in a non-optimized way. Human developers, in contrast, systematically navigated the inheritance tree, often employing a branch-and-bound style reasoning to cover broad categories first and then refine to specifics as needed.

Overall, this study highlighted that human experts leverage three key capabilities to excel in exception handling: \textit{Domain-Specific Knowledge}, \textit{Adaptive Error-Handling Strategies}, and \textit{Hierarchical Exception Reasoning}. These observations directly inform Seeker’s design: we incorporate an external knowledge base (CEE) to inject domain and technical-practice knowledge, use a multi-agent approach with specialized roles to allow adaptive and context-sensitive handling, and develop Deep-RAG to mimic hierarchical reasoning by efficiently searching the exception taxonomy.

\section{Preliminary}
\label{sec2.1} 

In this section, we explore how different prompting strategies affect an LLM’s exception handling performance. The intuition was that if we guide the LLM more like an expert would reason (using an intermediate “language” of hints and steps), the model might produce safer code. We defined four increasingly specific prompt styles (inspired by the expert strategies above):

\ding{182} \textbf{Coarse-Grained Reminding:} A minimal nudge to handle errors, e.g., \textit{“Remember to handle possible exceptions here.”} This prompt doesn’t specify which exceptions or how, just reminds the model to not ignore error handling.

\ding{183} \textbf{Fine-Grained Reminding:} Type-specific reminders, e.g., \textit{“This code involves database operations; consider catching a \texttt{SQLException} rather than a generic exception.”} This provides more concrete guidance on which exception to handle.

\ding{184} \textbf{Fine-Grained Inspiring:} Contextual scenarios with risk analysis, e.g., \textit{“If the network call fails (e.g., timeout or unreachable host), how should this be handled to maintain functionality?”} This style encourages the model to imagine failure scenarios and propose handling in context.

\ding{185} \textbf{Fine-Grained Guiding:} A structured, step-by-step directive that outlines how to handle the exception, possibly including inheritance considerations. In high level: \textit{“1) Identify all operations that can throw exceptions in this code. 2) For each, determine the most specific exception type (use known APIs/Java docs). 3) Draft try-catch blocks for those exceptions with appropriate fallback logic. 4) Make sure not to catch exceptions that shouldn’t be handled here (allow them to propagate if needed).”} This essentially walks the model through a mini-plan akin to how an expert would think.

\begin{figure*}[t]
 \centering
 \includegraphics[width=0.69\linewidth]{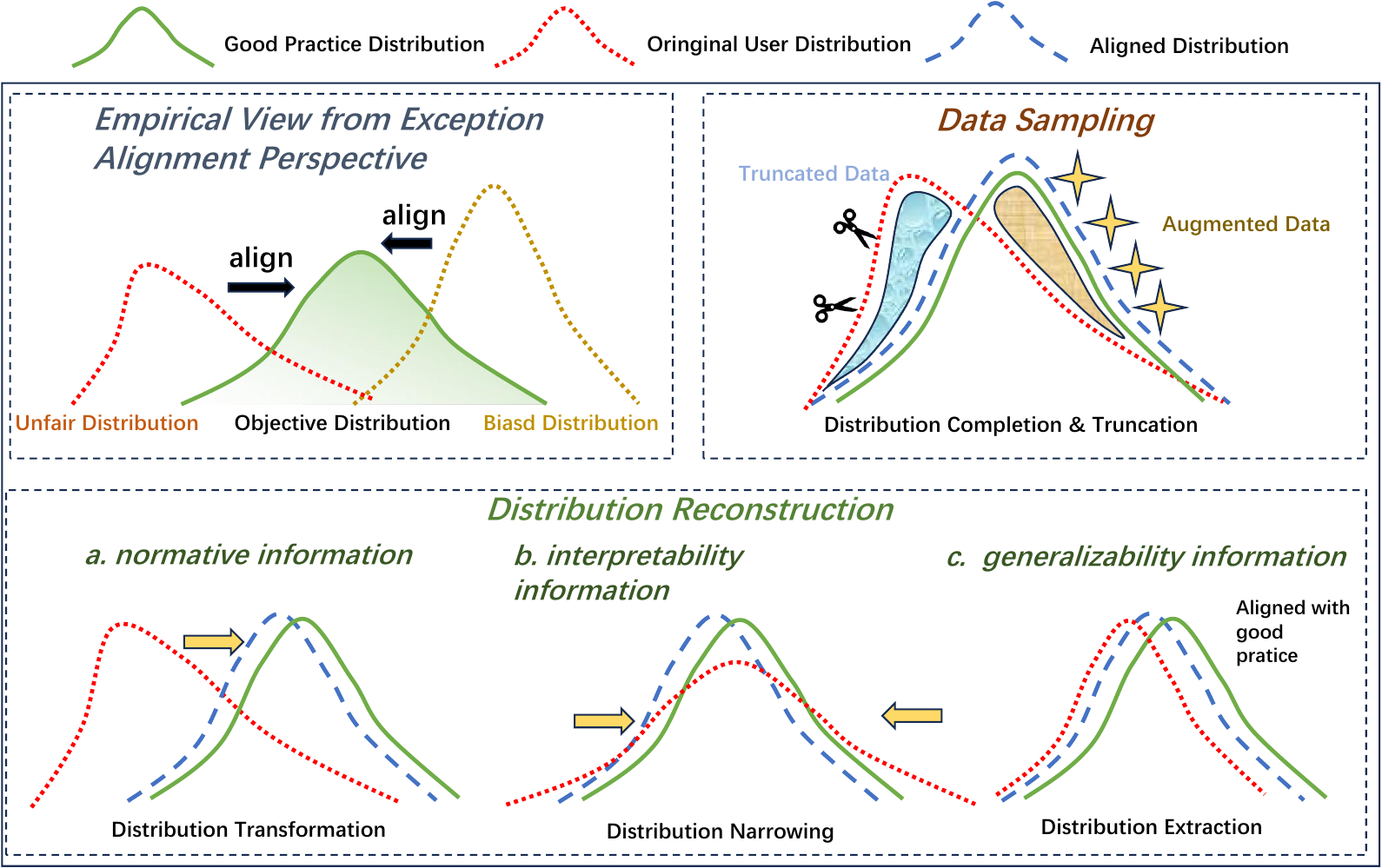}
 \caption{Aligning developers' exception handling from biased, user-oriented practices to industry-standard “good practice” distributions through iterative data refinement. Distribution truncation, augmentation, and reconstruction guide a progression from coarse-grained reminders to fine-grained, scenario-specific guidance—closing the gap between current human methods and stable, high-quality exception handling.}
 \label{fig3}
 \vskip -.1in
\end{figure*}

We tested these prompt styles on a subset of fragile code examples and measured the quality of the exception handling in the model’s output. Quality was evaluated via code review score by human experts (blindly, without knowing which prompt was used) on criteria including correctness of the handling, specificity of exceptions caught, and preservation of original functionality. Figure \ref{fig1.1} summarizes the results: moving from coarse to fine-grained guiding prompts, the LLM’s exception-handling performance steadily improved. With only coarse reminders, the model often inserted try-catch blocks but tended to use generic catches or simply print errors, doing little to resolve the problem. Fine-grained reminding (Prompt2) yielded more targeted catches (e.g., catching \texttt{NullPointerException} instead of \texttt{Exception} in one case, as suggested). The inspiring prompts further improved context-awareness; for instance, when asked to consider a network failure scenario, the model started to include retry logic or user notifications. The fine-grained guiding (Prompt4) had the most striking effect – the model’s outputs were markedly closer to human implementations, earning higher code review scores. It began to emulate a structured approach: identifying multiple exception types, handling each appropriately, and even refraining from catching exceptions that should be propagated (e.g., not catching \texttt{IOException} in a lower-level function that should let the caller handle it). This shows a clear mitigation effect: more detailed and structured prompts led to significantly better exception safety code.

We also qualitatively analyzed how human developers handle exceptions to derive these prompt strategies in Figure \ref{fig1.2} and \ref{fig3}. Human experts combine programming expertise, domain knowledge, and knowledge of exception hierarchies to craft robust solutions. For example, they might handle an \texttt{IOException} differently if it’s likely a \texttt{FileNotFoundException} (which can be handled by creating a file) versus a \texttt{SocketTimeoutException} (which might warrant a retry). They integrate adaptive strategies such as wrapping exceptions into custom ones in library code versus handling and logging in application code.

The prompting experiment validated that guiding the LLM with intermediate, structured reasoning steps dramatically improves exception handling output. This supports our premise that an Intermediate Representation (IR) – essentially a layer of communication that represents best practices and reasoning steps – can bridge the gap between the LLM’s default behavior and expert-level performance. However, manually writing elaborate prompts for each situation is not scalable. This motivates automating the process: Seeker’s agents effectively generate and process these intermediate “prompts” or representations internally. The IR in our framework manifests as natural language instructions, reasoning outputs, and the structured knowledge from CEE that together drive the LLM’s behavior in a controlled way.

\section{Common Exception Enumeration}
\label{sec3.2.2}

At the heart of Seeker’s knowledge integration is the Common Exception Enumeration (CEE), a structured repository of exception information and handling strategies. The goal of CEE is to encode, in a machine- and human-readable form, the collective wisdom on how to handle exceptions properly. CEE serves as the reference guide for our agents, especially the Predator and Ranker, when deciding \textit{what exceptions to handle and how to handle them.}

\subsection{Structure of CEE}
We organize CEE as a hierarchy mirroring the programming language’s exception class hierarchy. For example, in Java, exceptions form a tree under \texttt{Throwable} with branches such as \texttt{IOException} (and its subclasses), \texttt{RuntimeException} (and its subclasses like \texttt{NullPointerException}, \texttt{IndexOutOfBoundsException}, etc.), and so on. Each node in the CEE tree corresponds to an exception class (e.g., the node for \texttt{NullPointerException}). Each node stores the following key information:

\ding{182} \textbf{Basic Exception Info:} The name of the exception and its parent class (to situate it in the hierarchy). For checked exceptions, we note that as well, since handling requirements differ.

\ding{183} \textbf{Common Scenario:} A description of typical scenarios or conditions under which this exception is thrown. For instance, for \texttt{NullPointerException}, the scenario might be "calling a method on a \texttt{null} reference". For \texttt{FileNotFoundException}, it would be "attempting to open a file that does not exist or is inaccessible".

\ding{184} \textbf{Properties:} Any important attributes of this exception type. This can include whether it’s checked/unchecked, whether it indicates a serious error vs a minor one, or if it has known variants. It also include links to related exceptions (e.g., \texttt{SocketTimeoutException} is a kind of \texttt{IOException} with a specific cause).

\ding{185} \textbf{Recommended Handling Logic:} A concise guideline or template for how to handle this exception. This may reference best practices, such as "If possible, fallback to a default value or skip processing when catching \texttt{NumberFormatException}" or "Log the error and alert the user in the UI for exceptions related to user input". For some exceptions, the advice may be to not catch it at lower levels (for example, do not catch \texttt{OutOfMemoryError} or other serious \texttt{Error}s; those should propagate or crash).

Each exception node thus provides a capsule of knowledge: what it is, when it happens, and what to do (or not do) about it. Because this is structured, our agents can query this information programmatically. For example, the Predator agent might retrieve the "recommended handling logic" for a \texttt{SQLException} when it detects database code, and the Handler agent can then follow that logic in generating code. Here, we provide a sample.

\begin{tcolorbox}[breakable, title = Sample CEE Node]
\{\\
$\mbox{\ \ \ \ }$``name":$\mbox{\ }$``IOException",\\
$\mbox{\ \ \ \ }$``children":$\mbox{\ }$[...],\\
$\mbox{\ \ \ \ }$``info":$\mbox{\ }$\{\\
$\mbox{\ \ \ \ \ \ \ \ }$``definition":$\mbox{\ }$``IOException is a checked exception that is thrown when an input-output operation failed or interrupted. It's a general class of exceptions produced by failed or interrupted I/O operations.",\\
$\mbox{\ \ \ \ \ \ \ \ }$``reasons":$\mbox{\ }$``There are several reasons that could cause an IOException to be thrown. These include: File not found error, when the file required for the operation does not exist; Accessing a locked file, which another thread or process is currently using; The file system is read only and write operation is performed; Network connection closed prematurely; Lack of access rights.",\\
$\mbox{\ \ \ \ \ \ \ \ }$``dangerous\_operations":$\mbox{\ }$``Operations that could typically raise an IOException include: Reading from or writing to a file; Opening a non-existent file; Attempting to open a socket to a non-existent server; Trying to read from a connection after it's been closed; Trying to change the position of a file pointer beyond the size of the file.",\\
$\mbox{\ \ \ \ \ \ \ \ }$``sample\_code":$\mbox{\ }$``String fileName = 'nonexistentfile.txt'; \textbackslash n FileReader fileReader = new FileReader(fileName);",\\
$\mbox{\ \ \ \ \ \ \ \ }$``handle\_code":$\mbox{\ }$``String fileName = 'nonexistentfile.txt'; \textbackslash n try \{ \textbackslash n FileReader fileReader = new FileReader(fileName); \textbackslash n \} catch(IOException ex) \{ \textbackslash n    System.out.println('An error occurred while processing the file ' + fileName); \textbackslash n    ex.printStackTrace(); \textbackslash n \}",\\
$\mbox{\ \ \ \ \ \ \ \ }$``handle\_logic":$\mbox{\ }$``Try the codes attempting to establish connection with a file/stream/network, catch corresponding IOException and report it, output openpath is suggested."\\
$\mbox{\ \ \ \ }$\},\\
$\mbox{\ \ \ \ }$``scenario":$\mbox{\ }$``attempt to read from or write to a file/stream/network connection",\\
$\mbox{\ \ \ \ }$``property":$\mbox{\ }$``There might be an unexpected issue with accessing the file/stream/network due to reasons like the file not being found, the stream being closed, or the network connection being interrupted"\\
\}
\end{tcolorbox}

\subsection{Construction of CEE}
\begin{figure}[t]
  \centering
  \includegraphics[width=0.4\linewidth]{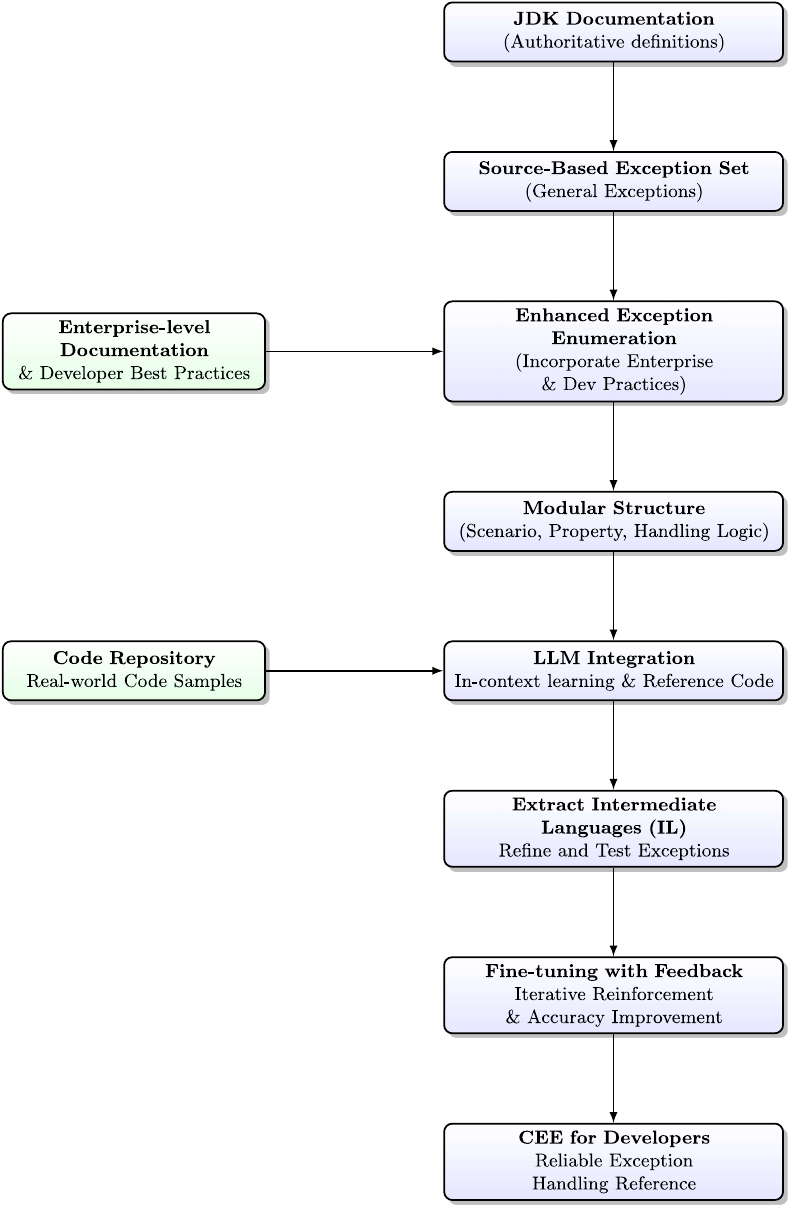}
  \caption{\textbf{An overview of the CEE construction process}. The diagram illustrates how authoritative documentation (JDK), enterprise-level best practices, and real-world code repositories are integrated and refined. Each exception node is enriched with Scenario, Property, and Handling Logic.  This framework is further optimized through LLM-based in-context learning and iterative fine-tuning, ultimately providing a reliable, structured reference (CEE) to enhance exception handling in generated code.}
  \label{fig:cee_schematic}
  \vskip -.2in
\end{figure}

Building CEE is a non-trivial task that we approached by merging multiple sources, as shown in Figure \ref{fig:cee_schematic}.

\ding{182} \textbf{Authoritative Documentation:} We started with official documentation (for Java, the JDK API docs) to list all standard exception classes and read their descriptions. This provided the skeleton of the hierarchy and basic info about each exception’s meaning and checked/unchecked status.

\ding{183} \textbf{Enterprise Practices:} We consulted technical-practice guides, style guides, expert-written blog posts and books on exception handling. These often contain rules of thumb (e.g., “Never catch Exception, instead catch specific subclasses” or “Close resources in a finally block or use try-with-resources”). We distilled such high-level guidelines and also more specific advice (how to handle specific exceptions in certain frameworks, if applicable).

\ding{184} \textbf{Real-World Repositories:} We mined a set of large open-source projects to see how exceptions are actually handled in practice. By analyzing commit messages and code, we identified common patterns: for example, many projects treat \texttt{IOException} in a similar way (logging and wrapping it into a runtime exception if they can’t recover), or handle \texttt{NumberFormatException} by providing default values. We also identified exceptions that are often ignored (which could indicate either they are truly ignorable or that developers commonly make a mistake by ignoring them). We fed these insights back into our knowledge base, refining the recommended handling. We also captured frequency – if a certain exception rarely occurs, our knowledge base still includes it, but our retrieval algorithm (Deep-RAG) will naturally focus on more likely ones unless context suggests otherwise.

CEE currently covers the full Java exception hierarchy (433 exception types) as well as custom exceptions that appeared in our dataset. This breadth ensures that even \textbf{long-tail exceptions} have an entry, so Seeker isn’t blind to any particular error. Of course, the depth of advice varies; common exceptions have rich guidance, while very rare ones might have basic default suggestions.

\subsection{Usage in Seeker}
We will discuss this part in detail in Section \ref{sec:method}. Simply put, CEE is used in two primary ways. First, the Detector/Predator agents leverage CEE to identify what exceptions might be relevant for a given code segment. For example, if Detector flags a code snippet with file I/O calls as fragile, Predator will query CEE for file-related exceptions (e.p. \texttt{FileNotFoundException}, \texttt{IOException}). It will use CEE’s scenario matching to see which exceptions in the knowledge base align with the operations in the code. Second, the Ranker/Handler agents use CEE to decide how to handle the exceptions that Predator chose. The Ranker pulls the recommended handling strategies for those exceptions from CEE and then decide which strategy fits best in context, and the Handler will then implement that strategy in code.
\begin{figure*}[t]
  \centering
  \includegraphics[width=0.99\linewidth]{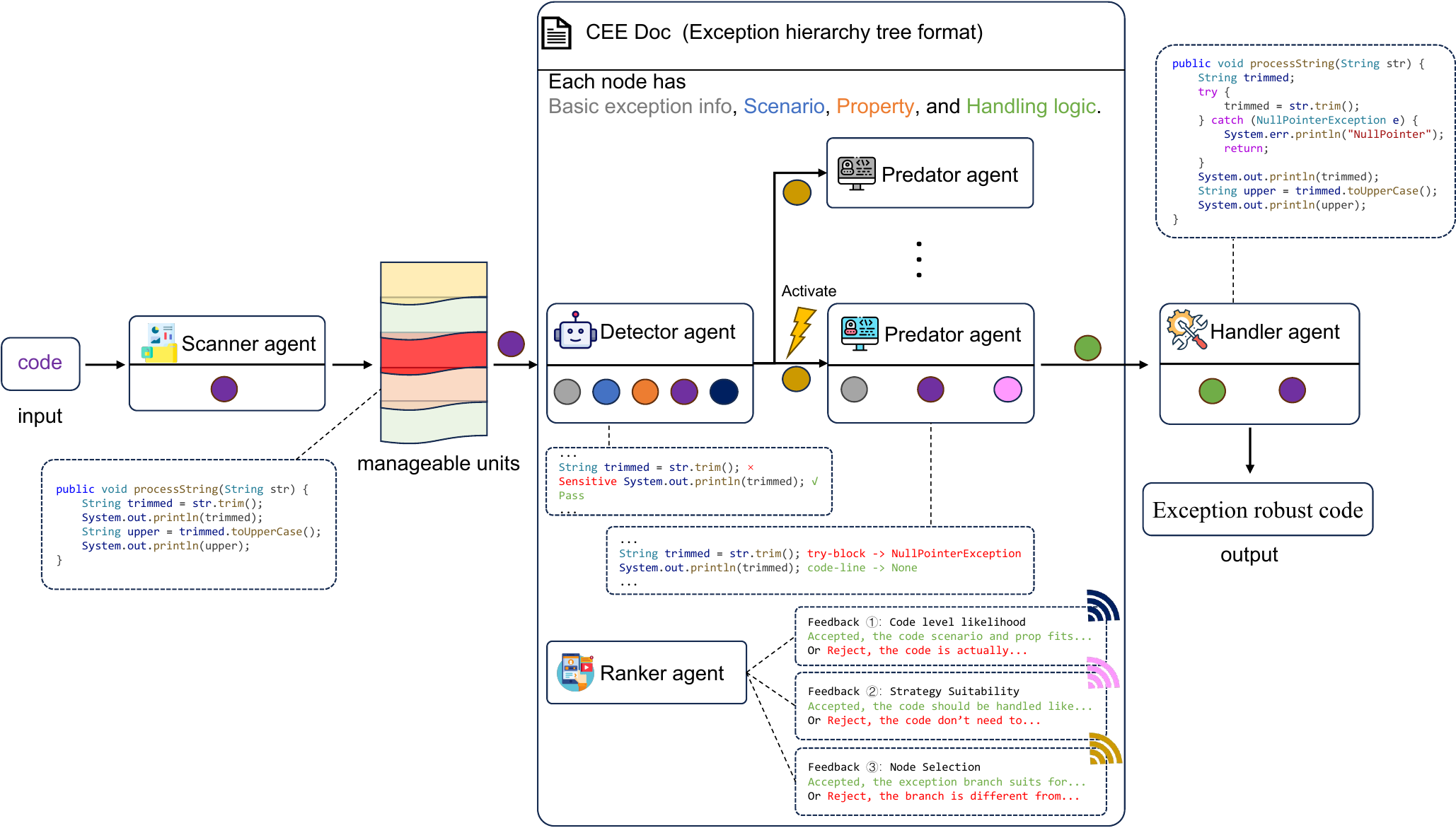}
  \caption{\textbf{Comprehensive Workflow of Seeker.} Seeker orchestrates the automated exception handling process through the seamless collaboration of five specialized agents: Scanner, Detector, Predator, Ranker, and Handler. The colored circles within the workflow illustrate the flow of information and interactions among the agents, highlighting how each component activates and contributes to the overall exception handling process. This integrated approach ensures that Seeker delivers highly reliable and maintainable exception handling solutions, significantly improving code robustness and developer productivity.}
  \label{fig4}
  \vskip -.1in
\end{figure*}
By incorporating CEE, Seeker benefits from an up-to-date, standardized set of exception handling practices. This helps prevent the model from regressing into bad habits (like catching broad exceptions or ignoring errors), as the knowledge base consistently pushes it toward what an expert would do. It also makes the system more interpretable: one can inspect which CEE entries were used in a decision, effectively explaining why a certain handling was inserted (e.g., "we handled \texttt{SQLException} because CEE indicates it's a common, important exception for DB operations, and we used a fallback query based on the CEE’s recommendation").

Finally, CEE is designed to be \textbf{extensible and community-maintainable}. Developers could contribute new patterns or modifications to the knowledge base as new frameworks or technical practices emerge. In essence, CEE bridges human expertise with automated generation, ensuring that as best practices evolve, the LLM’s behavior can be updated without retraining – simply by updating the knowledge entries, following the new trend of dynamic \cite{multicollinearity}.

\section{Methodology}
\label{sec:method}
\subsection{Seeker Multi-Agent Framework}
The Seeker framework comprises five specialized agents that collaborate to transform input code into an exception safety output. Figure \ref{fig4} illustrates the overall workflow. By decomposing the task, each agent can focus on a specific sub-problem, making the overall process more manageable and interpretable. We describe each agent and its role below:

\begin{itemize}
\item \textbf{Scanner:} The Scanner agent is responsible for parsing and partitioning the code into manageable units for analysis. Given an input (which could be a function, a code file, or a snippet), the Scanner divides it into logical segments, such as individual functions, code blocks, or dependency chains. The idea is to limit the scope so subsequent agents can focus on one segment at a time, and nothing important is missed in large files. The Scanner also performs a preliminary scan for obvious red flags (like the presence of try/catch keywords or lack thereof, use of throws declarations, etc.). Essentially, it outputs a list of fragile code candidates: code units that should be examined for potential exception issues. For instance, in a code snippet:
\begin{tcolorbox}[breakable, title = ]
\textcolor{purple}{public void} \textcolor{blue}{processString}(String str) \{\\
$\mbox{\ \ \ \ }$\textcolor{orange}{String trimmed} = str.trim();\\
$\mbox{\ \ \ \ }$System.out.println(trimmed);\\
$\mbox{\ \ \ \ }$\textcolor{orange}{String upper} = trimmed.toUpperCase();\\
$\mbox{\ \ \ \ }$System.out.println(upper);\\
\}
\end{tcolorbox}
the Scanner will isolate the body of \texttt{processString} as a unit and note that \texttt{str.trim()} could be risky (since \texttt{str} might be null).
\item \textbf{Detector:} The Detector agent takes the units from the Scanner and identifies which segments are truly fragile (i.e., likely to throw or propagate exceptions that are not being handled). It uses heuristics and possibly model predictions to find “sensitive” code lines. For example, it knows common operations that can throw exceptions (file I/O, network calls, type parsing, etc.), and it checks if those are unprotected. In the example above, Detector would flag the line \texttt{String trimmed = str.trim();} as fragile because \texttt{trim()} can throw a \texttt{NullPointerException} if \texttt{str} is null. It marks the entire \texttt{processString} method as needing exception handling, classifying it as a detection. Essentially, Detector outputs a refined list of segments annotated with specific spots/queries that need exception coverage. In our framework, we formalize these as queries or signals (e.g., “What if \texttt{str} is null here?”).
\item \textbf{Predator:} The Predator agent “hunts down” possible exceptions for each fragile segment identified. Using the queries from Detector, Predator interacts with the CEE knowledge base to enumerate which exceptions could occur. Continuing the example, Predator knows from CEE that \texttt{trim()} on a null string triggers a \texttt{NullPointerException}. It might also consider other exceptions (though \texttt{trim()} likely only that one in this simple case). Predator essentially produces a set of candidate exceptions (and their corresponding handling needs) for that code segment. In more complex code, Predator lists multiple exceptions; for instance, if the code opens a file and reads from it, Predator would list \texttt{FileNotFoundException}, \texttt{IOException}, etc. Predator uses the Deep-RAG algorithm (detailed in Section \ref{apx1.1}) to efficiently retrieve relevant exceptions from CEE without combing through the entire hierarchy blindly. The output of Predator is a tentative mapping: \textit{“In this code segment, exceptions E1, E2, ... En are likely and need handling.”} For each such exception, Predator also fetch the recommended handling logic outline from CEE for the Ranker’s benefit.
\item \textbf{Ranker:} If multiple exception handling options exist, the Ranker agent evaluates and selects the most appropriate handling strategy for the current context. There are two dimensions for this. First, which exceptions to handle explicitly – sometimes Predator might list an exception that, on second thought, doesn’t need a local catch (maybe it should bubble up). Ranker decides if an exception should indeed be caught here or if it’s better left to a higher level. It uses cues like the severity (from CEE properties) and the context (is this a library function or a top-level call?). Second, for each exception that will be handled, how to handle it. If CEE provides multiple strategies (for example, one strategy might be “retry once then fail”, another might be “log and continue with default”), Ranker evaluates which fits the scenario. The evaluation considers the function’s role – e.g., if it’s a user-facing function, logging and showing an error is better; if it’s a library, propagating an exception is better. Ranker simulates outcomes using LLM’s reasoning. In our design, we prompt LLM (in Ranker persona) to decide: \textit{“Given the knowledge base suggestions and the code context, which handling approach ensures safety and minimal disruption?”} The outcome of Ranker is a concrete plan for the Handler: which exceptions to catch and what each catch block should do.
\item \textbf{Handler:} The Handler agent is the final actor that generates the actual code modifications to implement the chosen exception handling plan. It takes the original code and inserts or modifies it to include try-catch blocks, resource-finally blocks, or patterns as decided. The Handler ensures the inserted code follows the language syntax and style, and importantly, that it doesn’t break the original functionality. For example, in \texttt{processString}, the Handler would wrap \texttt{str.trim()} in a try-catch, and then proceed with the rest of the logic (\texttt{System.out.println(trimmed); ...}). This matches what an expert might do: catch \texttt{NullPointerException} and handle it (here, logging and early return to avoid using a null). The Handler’s output is the fully transformed code with robust exception handling.

\end{itemize}
\begin{algorithm*}[t]
\small
\caption{Seeker Framework}
\label{ag11}
\KwIn{Codebase $C$}
\KwOut{Optimized code $C'$ with robust exception handling}
% Step 1: Segmentation
Segment the codebase $C$ into manageable units $U = \{u_1, u_2, \dots, u_N\}$\;
% Define criteria for manageable units
\ForEach{code segment $u_i$ in $C$}{
    \If{(length of $u_i$ is within predefined limit) \textbf{and} (function nesting level is low) \textbf{and} (logical flow is clear)}{
        Add $u_i$ to $U$\;
    }
}
Initialize optimized units $U' = \{\}$\;
\ForEach{unit $u_i$ in $U$}{
    % Detection Phase
    \tcp{Detection Phase}
    Initialize potential exception set $E_i = \{\}$\;
    Use the \textbf{Detector} agent to analyze unit $u_i$\;
    \textbf{In parallel do} \{
        \tcp{Static Analysis}
        Generate control flow graph $CFG_i$ and exception propagation graph $EPG_i$ for $u_i$\;
        Identify sensitive code segments $S_{i}^{\text{static}} = \{s_{i1}^{\text{static}}, s_{i2}^{\text{static}}, \dots\}$ in $u_i$\;
        \tcp{Scenario and Property Matching}
        Perform scenario and property matching on $u_i$\;
        Identify sensitive code segments $S_{i}^{\text{match}} = \{s_{i1}^{\text{match}}, s_{i2}^{\text{match}}, \dots\}$ in $u_i$\;
    \}
    Combine sensitive code segments: $S_i = S_{i}^{\text{static}} \cup S_{i}^{\text{match}}$\;
    \ForEach{segment $s_{ij}$ in $S_i$}{
        Detect potential exception branches $E_{bij}$ in $s_{ij}$\;
        $E_{bi} \leftarrow E_{bi} \cup E_{bij}$\;
    }

    % Retrieval Phase
    \tcp{Retrieval Phase}
    Use the \textbf{Predator} agent to retrieve fragile code and try-catch blocks\;
    Summarize unit $u_i$ at the function level to obtain code summary $F_i$\;
    Perform Deep-RAG using $F_i$ and exception branches $E_{bi}$, get exception nodes $E_{ni}$\;
    Mapping relevant exception handling strategies $H_i = \{h_{i1}, h_{i2}, \dots\}$ from CEE\;
    % Ranking Phase
    \tcp{Ranking Phase}
    Use the \textbf{Ranker} agent to assign grades to exceptions in $E_{ni}$\;
    \ForEach{exception $e_{ik}$ in $E_{ni}$}{
        Calculate exception likelihood score $l_{ik}$ based on $e_{ik}$ attribute and impact\;
        Calculate suitability score $u_{ik}$ of handling strategy $h_{ik}$\;
        Compute overall grade $g_{ik} = \alpha \cdot l_{ik} + \beta \cdot u_{ik}$\;
    }
    Rank exceptions in $E_{ni}$ based on grades $g_{ik}$ in descending order to get ranked list $E_{ni}'$\;
    % Handling Phase
    \tcp{Handling Phase}
    Use the \textbf{Handler} agent to generate optimized code $u_i'$\;
    \ForEach{exception $e_{ik}$ of $E_{ni}'$ if $g_{ik} > \gamma$}{
        Mapping handling strategy $h_{ik}$ from $H_i$\;
        Apply $h_{ik}$ to code segment(s) related to $e_{ik}$ in $u_i$\;
    }
    $U' \leftarrow U' \cup \{u_i'\}$\;
}
% Step 3: Combine optimized units
Combine optimized units $U'$ to produce the final optimized code $C'$\;
\end{algorithm*}

The Intermediate Representation (IR) aspect is evident in the interactions among these agents. Instead of directly asking the LLM “write safe code,” we guide it through intermediate steps. Detector might output a note to Predator: “Possible null dereference at line 3. Query: what exception arises if str is null?”. Predator outputs to Ranker: “Exception NullPointerException likely at line 3; recommended strategies: [1] handle locally by logging and returning default, [2] let it propagate.”. This back-and-forth is the IR – a layer of reasoning and explanation that is not part of the final code but crucial in producing it. It is human-interpretable, meaning one could read the agents’ dialogue and understand the rationale for the changes (which aids debugging and trust in the system).
\subsection{Deep Retrieval-Augmented Generation Algorithm}
\label{apx1.1}
Consider Java’s exception hierarchy – with over 400 exception types, 62 distinct branches, and up to 5 levels of depth. Handling exceptions effectively often means dealing with long inheritance chains and a vast space of possible error types. A naive approach might try to enumerate every exception or search through all of CEE for each fragile code – which would be inefficient and could overwhelm the LLM with irrelevant information. Our Deep Retrieval-Augmented Generation (Deep-RAG) algorithm is designed to tackle this by combining structured retrieval with the generative reasoning of the LLM, pruning the search space early and focus on the most likely branches. Deep-RAG works in two phases:

\ding{182} \textbf{Branch Identification via Scenario Labels:} We pre-process the exception hierarchy by assigning “development scenario” labels to top-level branches or clusters of exceptions. These labels are essentially categories or contexts. For example, one branch label might be "File/IO Operations" (covering \texttt{IOException} and its subclasses), another "Network/Communication Errors" (covering exceptions like \texttt{SocketException}, \texttt{TimeoutException}), another "Null/Type Errors" (covering \texttt{NullPointerException}, \texttt{ClassCastException}, etc.), "Security/Access" (for \texttt{SecurityException}, etc.), "Concurrency" (for \texttt{InterruptedException}, \texttt{ConcurrentModificationException}), and so on. We derive these categories from CEE and the kinds of operations typically leading to those exceptions.

When the Predator agent examines a code segment, it first summarizes the unit context (what operations or APIs are used here). For example, if the code opens a URL and reads data, the summary might be "network IO operation". Deep-RAG then maps this summary to one or more scenario labels (e.g., "Network/Communication Errors" and "File/IO" if file is involved too). Essentially, it reasons which branches of the exception tree are relevant. This uses a few-shot learning model: we gave examples to the LLM of code scenarios and which labels they correspond to, enabling it to generalize. By doing this, we reduce the search space dramatically – out of dozens of branches, 2–3 are selected for deeper retrieval. This step is akin to an expert quickly saying: "This code deals with a network call, so likely network-related exceptions might happen (like timeouts, unreachable host, etc.), no need to worry about, say, database exceptions."

\ding{183} \textbf{Focused Exception Retrieval and Generation:} Given the relevant branches, Deep-RAG traverses those parts of CEE hierarchy to retrieve specific exception types that match more closely the operations in code. For each branch, ask the model which exception property in this branch could be triggered by this code. For instance, within "Network/Communication Errors", the algorithm looks at IOException branch and specifically its network-related subclasses (e.p. \texttt{SocketTimeoutException}, \texttt{UnknownHostException}). It uses a combination of rules to verify if those are plausible (for example, if the code is opening a socket, a timeout is plausible; if the code doesn’t do any DNS resolution, \texttt{UnknownHostException} will not apply). The output of this step is a set of exception nodes (specific exceptions) denoted as likely relevant.

Alongside retrieving exception names, Deep-RAG pulls the associated handling strategies from CEE for those exceptions. This yields a package of knowledge. LLM uses this retrieved knowledge to propose how to handle these exceptions in context. 

The generative power of LLM is used to adapt the generic knowledge to the specific code. For example, CEE might say "for \texttt{SocketTimeoutException}, you might retry the operation", and LLM incorporates that into the actual code’s logic, suggesting a loop around the network call with a retry.

\begin{algorithm*}[t]
\caption{Deep Retrieval-Augmented Generation (Deep-RAG)}
\KwIn{Knowledge hierarchy tree $T$, unit summary $F_i$, detected queries $Q_i$, environment context $Env$}
\KwOut{Relevant information retrievals $R_i$}
Initialize relevant knowledge branches set $B = \{\}$\;
Assign knowledge scenario labels $L = \{l_1, l_2, \dots\}$ to branches of $T$\;
% Identify relevant knowledge branches
\ForEach{query $q_{ik}$ in $Q_i$}{
    Identify branches $B_{ik}$ in $T$ related to $q_{ik}$ based on labels $L$\;
    $B \leftarrow B \cup B_{ik}$\;
}
% Verification Step
\ForEach{branch $b_m$ in $B$}{
    \tcp{Verification Step}
    Select few-sample document examples $X_m = \{x_{m1}, x_{m2}, \dots\}$ associated with branch $b_m$\;
    \ForEach{example $x_{mj}$ in $X_m$}{
        Perform query matching to obtain pass rate $p_{mj}$ and capture accuracy $a_{mj}$\;
        \If{$p_{mj}$ or $a_{mj}$ below threshold $\theta$}{
            Record failure pattern $fp_{mj}$ based on $Env$\;
            Update environment context $Env$ with $fp_{mj}$\;
        }
    }
    Compute average pass rate $\bar{p}_m$ and accuracy $\bar{a}_m$ for branch $b_m$\;
    \If{$\bar{p}_m$ or $\bar{a}_m$ below threshold $\theta$}{
        Fine-tune labels $L$ for branch $b_m$ based on aggregated feedback from $Env$\;
    }
}
% Node Evaluation
Initialize information retrievals set $R_i = \{\}$\;
\ForEach{branch $b_m$ in $B$}{
    Select depth level $D$ for node evaluation\;
    \For{$d = 1$ to $D$}{
        \ForEach{node $n_{ml}$ at depth $d$ in branch $b_m$}{
            Evaluate relevance score $r_{ml}$ to summary $F_i$ and queries $Q_i$\;
            \If{$r_{ml} > \delta$}{
                Retrieve information $r_{ml}$ from knowledge base\;
                $R_i \leftarrow R_i \cup \{r_{ml}\}$\;
            }
        }
    }
}
\end{algorithm*}
\begin{table}[ht]
\centering
\caption{Computation Time Before and After Parallelization}
\label{table:computation_cost}
\resizebox{0.8\columnwidth}{!}{
\begin{tabular}{lcc}
\toprule
\textbf{Processing Method} & \textbf{Average Time per Code File (s)} & \textbf{Speedup Factor} \\
\midrule
Sequential Processing      & 291.0 & 1x \\
Parallel Processing (Seeker) & 19.4  & 15x \\
\bottomrule
\end{tabular}
}
\end{table}
In essence, Deep-RAG selectively activates portions of the knowledge base relevant to the code at hand, rather than dumping everything. This leads to both efficiency gains and accuracy improvements. The 93\% reduction in computational overhead we reported in Table \ref{table:computation_cost} comes from not having to consider irrelevant branches and exceptions. Another advantage is generalizability: by focusing on scenario labels and branches, Deep-RAG can be extended to other languages or domains relatively easily. For a new language, one would label that language’s exception categories (e.g., for Python - OS errors, value errors, etc.), and the same approach holds. The algorithm’s label-to-branch mapping is learned with a few examples, meaning even if the domain changes (say, we apply it to web application errors, where scenarios are "HTTP errors", "database errors"), the mechanism is the same – just the labels and knowledge base content change. This demonstrates a flexible, agent-based interaction model that allows the system to adapt without redesign.

\section{Study Design}
\label{experiments}
In this section, we design a large-scale study to assess \textbf{Seeker} by answering five research questions:

\begin{itemize}
    \item RQ1: How does Seeker’s exception-handling performance compare to state-of-the-art baselines?
    \item RQ2: What is the contribution of each agent in Seeker’s multi-agent framework (ablation study)?
    \item RQ3: How does the choice of underlying LLM (open-source vs. closed-source) affect Seeker’s performance?
    \item RQ4: How critical is the integration of CEE knowledge base to Seeker’s success?
    \item RQ5: Can Seeker generalize its benefits to other tasks or benchmarks beyond the primary dataset (e.g., real bug fix tasks and general code generation challenges)?
\end{itemize}

We also describe the details of the study, including datasets, baselines, evaluation metrics and underlying models.

\subsection{Tasks \& Datasets}
We assembled a primary evaluation dataset of 15 real-world Java projects (sourced from GitHub, 2019–2024) containing a total of 750 fragile code snippets that need improved exception handling. To ensure the quality and representativeness of the dataset, we carefully selected projects that are both active and large in scale. Following the previous work, we applied stringent selection criteria, including the number of stars, forks, and exception handling repair suggestions in the project, to ensure that the dataset comprehensively covers the exception handling practices of modern open-source projects \cite{baseline3}. These projects span various domains (utilities, web backends, data processing libraries) to ensure diversity. 

\begin{table*}[th]
\centering
\caption{\textbf{The Excerpt Data source.} We quantify the quality of datasets in the context of code generation and exception handling using multiple dimensions, encompassing project popularity, community engagement, codebase quality, security posture, documentation integrity and dynamic maintenance.}
\label{table:source}
\begin{tabular}{lcccccc}
\toprule
\textbf{Repo} & \textbf{Commits} & \textbf{Stars} & \textbf{Forks} & \textbf{Issue Fix} & \textbf{Doc} & \textbf{Under Maintenance}\\
\midrule
Anki-Android       & 18410  & 8500  & 2200  & 262  & Y & Y \\
AntennaPod      & 6197  & 6300  & 1400  & 295  & Y & Y  \\
connectbot           & 1845  & 2480 & 629  & 321  & N/A & Y \\
FairEmail             & 30259  & 3073  & 640  & N/A  & Y & Y  \\
FBReaderJ        & 7159  & 1832  & 802  & 248  & Y & N/A \\
FP2-Launcher        & 1179  & 25  & 2  & 16 & Y & N/A \\
NewsBlur          & 19603  & 6800  & 995  & 158 & Y & Y \\
Launcher3        & 2932 & 91 & 642 & 2 & N/A & Y \\
Lawnchair-V1      & 4400  & 93  & 43  & 394  & Y & Y \\
MozStumbler      & 1727  & 619  & 212  &  203 & Y & N/A \\
\bottomrule
\end{tabular}
\end{table*}

To provide a holistic assessment, we propose a Composite Quality Metric (CQM) that aggregates these dimensions into a single quantitative indicator. Open source code repositories that perform well under this metric enter our semi-automated review process to screen high-quality exception handling blocks for few-shot, CEE building, or testing. The code snippets were selected in two ways: (1) by mining historical commits for instances where developers later added exception handling (indicating the original was fragile), and (2) by static analysis plus manual review to find code that likely lacks necessary try-catch blocks (following criteria similar to our preliminary study, e.g., calls to methods that throw checked exceptions with no catch). 

To avoid data leakage \cite{donggeneralization}, we also performed a round of variations on the test set. Considering that our method does not directly rely on data but fully utilizes LLM's ability to understand and reason about code, the evaluation results are consistent with our predictions, and the impact of data leakage on the credibility of our method is negligible.

This dataset represents realistic scenarios where exception handling is deficient or could be improved. We use this as the main testbed for RQ1–RQ4. Additionally, for the generalization analysis, we used:

\begin{itemize}
    \item \textbf{SWE-bench \cite{swebench}:} Focusing on real-world issue resolution tasks.
    \item \textbf{CoderEval \cite{codereval}:} A benchmark for pragmatic code generation tasks, where the goal is to generate functionally correct Java code that often involves using custom libraries or handling multiple components. We specifically find that exception handling could influence success (e.g., tasks requiring reading input where errors might occur).
\end{itemize}

\subsection{Baselines}
We compare \textbf{Seeker} with the following methods:
\begin{itemize}
    \item \textbf{General Prompting}: A straightforward baseline where we prompt a strong LLM (GPT-4o in our case) with the raw code and a generic instruction to "add proper exception handling". This represents the naive approach without our structured framework.
    \item \textbf{Web-search (RAG)}: A retrieval-augmented baseline that uses web search (StackOverflow, documentation) for each snippet to find relevant info, then feeds that to the LLM to guide code generation. This simulates how a developer might manually search for how to handle an error and how an RAG approach (without our specialized structure) might perform.
    \item \textbf{KPC} \citep{kpc}: The state-of-the-art Knowledge-driven Prompt Chaining method for exception handling code generation which chains prompts to the LLM including knowledge from API docs.
    \item \textbf{FuzzyCatch} \citep{baseline3}: A classical tool for recommending exception handling code based on fuzzy logic.
    \item \textbf{Nexgen} \citep{baseline1}: A neural network pretraining approach for automated exception handling in Java.
\end{itemize}
These baselines cover a range from naive to deep learning approaches. It is worth noting that a few works are not considered due to the limitation of their code granularity, which is not specialized and cannot be reasonably applied to our test cases. They are covered as much as possible in Section \ref{rw222}.

\subsection{Evaluation Metrics}
\label{sec:setup:metrics}
To holistically evaluate \textbf{Seeker}'s effectiveness in enhancing code robustness, we employ six metrics spanning three critical dimensions:  

\begin{enumerate}
    \item \textbf{Detection Efficacy}  
    \begin{itemize}
        \item \textit{Coverage (COV)}: Percentage of actual fragile code segments detected, measuring recall:  
        
        For the codebase under test, let \( S = \{ s_1, s_2, \dots, s_N \} \) be the set of actual sensitive code segments.

        Let \( D = \{ d_1, d_2, \dots, d_M \} \) be the set of detected sensitive code segments.

        \[
        I_{\text{detected}}(s_i) = \begin{cases}
        1, & \text{if } \exists d_j \in D \text{ such that } d_j = s_i \\
        0, & \text{otherwise}
        \end{cases}
        \]

    \[
    \text{COV} = \frac{\sum_{i=1}^N I_{\text{detected}}(s_i)}{N} \times 100\%
    \]
        
        \item \textit{Coverage Pass (COV-P)}: Precision-adjusted detection of try-blocks by the Predator agent, penalizing over-detection (FP) while rewarding alignment with ground truth:  

        For all code segments, let \( T = \{ t_1, t_2, \dots, t_P \} \) be the set of actual code regions that should be enclosed in try-catch blocks (actual try-blocks).

        Let \( \hat{T} = \{ \hat{t}_1, \hat{t}_2, \dots, \hat{t}_Q \} \) be the set of code regions detected by the \textbf{Predator} agent as requiring try-catch blocks (detected try-blocks).

        \[
    I_{\text{correct}}(\hat{t}_j) = \begin{cases}
    1, & \text{if } \hat{t}_j \in T \\
    0, & \text{otherwise}
    \end{cases}
    \]
    Therefore, $TP = \sum_{j=1}^Q I_{\text{correct}}(\hat{t}_j)$, $FP = Q - TP$, $FN = P - TP$

    \[
    \text{COV-P} = \frac{\text{TP}}{P + \text{FP}} \times 100\%
    \]
        
        %\[
        %\text{COV-P} = \frac{\text{Correct Try-Blocks}}{\text{Detections} + \text{Missed %Actuals}} \times 100\%
        %\]
    \end{itemize}

    \item \textbf{Exception Handling Accuracy}  
    \begin{itemize}
        \item \textit{Type Accuracy (ACC)}: Correctness of identified exception types (incl. subclass relationships):  
        
        For each element in the union of actual try-blocks and detected try-blocks, let \( E = \{ e_1, e_2, \dots, e_R \} \) be the set of actual exception types that should be handled.

    Let \( \hat{E} = \{ \hat{e}_1, \hat{e}_2, \dots, \hat{e}_S \} \) be the set of exception types identified by the \textbf{Predator} agent.

    \[
    I_{\text{correct}}(\hat{e}_j) = \begin{cases}
    1, & \text{if } \hat{e}_j = e_i \\
    1, & \text{if } \hat{e}_j\text{ is a subclass of } e_i\\
    0, & \text{otherwise}
    \end{cases}
    \]

    \[
    \text{ACC} = \frac{\sum_{j=1}^S I_{\text{correct}}(\hat{e}_j)}{S} \times 100\%
    \]
        
        \item \textit{Edit Similarity (ES)}: Structural fidelity of generated try-catch blocks vs. human solutions:  

        For each element in the union of actual try-catch-blocks and detected try-catch-blocks, let \( G \) be the generated try-catch code, and \( A \) be the actual try-catch code.

        \[
    \text{ES} = 1 - \frac{\text{LevenshteinDistance}(G, A)}{\max(|G|, |A|)}
    \]
    
        %\[
        %\text{ES} = 1 - \frac{\text{Levenshtein Distance}}{\max(\text{Generated}, %\text{Reference})}
        %\]
    \end{itemize}

    \item \textbf{Code Quality}  
    \begin{itemize}
        \item \textit{Automated Code Review Score (ACRS)}: Weighted compliance with 32 enterprise standards (security, maintainability, etc.), where $w_i$ is the weight assigned to the $i$-th code quality rule reflecting its importance, $s_i$ is the score for the $i$-th rule, defined as $s_i = \frac{q_i}{Q_i}$, where $q_i$ is the raw score for the $i$-th rule based on the specific quality measure (e.g., code readability, efficiency, etc.), $Q_i$ is the maximum possible score for the $i$-th rule which ensures that $s_i$ is normalized to the range $[0, 1]$:  
        
        \begin{equation}
    \text{ACRS} = \frac{\sum_{i=1}^{32} w_i s_i}{\sum_{i=1}^{32} w_i} \times 100\%
    \end{equation}
        
        \item \textit{Code Review Score (CRS)}: LLM-based assessment (GPT-o1) of exception handling practices:  

        For each exception-occur functions, let \( N_{\text{good}} \) be the number of generated try-catch blocks evaluated as \emph{good}, and \( N_{\text{total}} \) be the total number of try-catch blocks evaluated.
        
        \[
    \text{CRS} = \frac{N_{\text{good}}}{N_{\text{total}}} \times 100\%
    \]
    \end{itemize}
\end{enumerate}

\textbf{Rationale \& Novelty}: While COV/ACC address traditional detection tasks, COV-P introduces inheritance-aware precision critical for Java's deep exception hierarchies. ES complements syntactic similarity with semantic validity through ACRS/CRS - a dual-assessment strategy overcoming limitations of single-metric evaluations in prior work \citep{kpc, baseline3}. Our metrics are explicitly designed to measure:  

- \textit{Generalizability}: Via cross-language CRS validation  

- \textit{Interpretability}: Through ACRS' rule-based breakdown  

- \textit{Robustness}: By combining detection (COV) with handling quality (ES/CRS)  

The metric suite enables granular analysis of \textbf{Seeker}'s components while aligning with software engineering quality standards [ISO-25010].

\subsection{Underlying Models}
We use GPT-4o \cite{gpt4o} as default underlying model. We also use different open-source (e.g. Code Llama-34B \citep{codellama}, WizardCoder-34B \citep{wizardcoder}, Vicuna-13B \citep{vicuna}) and closed-source(e.g. Claude-2 \citep{claude}, GPT-3-davinci \citep{gpt3}, GPT-3.5-turbo \citep{gpt3.5}, GPT-4-turbo \citep{gpt4} LLMs as underlying model to further analyze models' ability for exception handling in RQ3.

\section{Results Analysis}
\label{sec:result}

\subsection{RQ1: Performance Comparison with Baselines}
We compare the performance of \textbf{Seeker} against baseline methods on the exception handling code generation task. The results are summarized in Table~\ref{table:results}.

\begin{table}[ht]
\centering
\caption{Comparison of Exception Handling Code Generation Methods}
\label{table:results}
\resizebox{0.8\columnwidth}{!}{  % Resize to fit one column width
\begin{tabular}{lcccccc}
\toprule
\textbf{Method} & \textbf{ACRS} & \textbf{COV (\%)} & \textbf{COV-P (\%)} & \textbf{ACC (\%)} & \textbf{ES} & \textbf{CRS (\%)} \\
\midrule
General Prompting       & 0.21 & 13 & 9 & 8 & 0.15 & 24 \\
Web-search          & 0.35 & 35 & 31 & 29 & 0.24 & 31 \\
KPC          & 0.26 & 14 & 11 & 8 & 0.17 & 27 \\
FuzzyCatch & 0.47 & 52 & 50 & 43 & 0.36 & 48 \\
Nexgen & 0.45 & 56 & 49 & 42 & 0.41 & 52 \\
\textbf{Seeker (Ours)}      & \textbf{0.85} & \textbf{91} & \textbf{81} & \textbf{79} & \textbf{0.64} & \textbf{92} \\
\bottomrule
\end{tabular}
}
\end{table}

Seeker achieves the best performance on all metrics by a significant margin, demonstrating its effectiveness:

\begin{itemize}
    \item Average Code Review Score (ACRS): Seeker’s ACRS is 0.85, far higher than the baselines (second, Nexgen, is 0.45). This indicates an overall superior code quality after applying Seeker. Essentially, on average, the quality of Seeker’s output was rated 85\% of an ideal expert solution, whereas others were around 21–47\%. This gap underscores how the multi-agent guided approach produces more thorough and correct modifications.
    \item Coverage (COV) and Coverage-Pass (COV-P): Seeker covers 91\% of the fragile code segments, meaning it detects and handles almost all the places it should. Baselines like FuzzyCatch and Nexgen achieved around 52–56\% coverage, missing nearly half of the problematic spots. Moreover, Seeker’s COV-P is 81\%, indicating most of its detected catches are correct and necessary. The drop from 91 to 81 suggests a small number of over-catches, but still far better than others (baselines had COV-P around 9–50\% only). In practical terms, Seeker is exceptional at precise fragile blocks catching.
    \item Accuracy (ACC): Seeker reached 79\% accuracy in identifying the correct exception types to catch, which is nearly double the 42–43\% of the best baseline. It means in a majority of cases, Seeker’s catches were of the specific exceptions that a software would expect, including correctly recognizing subclass relationships. The knowledge-driven nature (via CEE and Deep-RAG) clearly helps here – Seeker doesn’t randomly catch exceptions; it picks the right ones with high precision.
    \item  Edit Similarity (ES): With ES = 0.64, Seeker’s changes were very close to human fixes. The second, Nexgen, had 0.41. Reviewers noted that code produced by Seeker often mirrored best-practice patterns that humans use (like exact error messages, structures of try-catch), hence the high similarity. This metric highlights not just correctness but the style and intent alignment with human solutions.
    \item Code Review Score (CRS): Seeker scored 92\% on code review evaluations. Essentially nearly all its output was deemed acceptable or good in a code review context, while baselines lagged (31–52\%). Notably, the general prompting baseline scored extremely low across the board (CRS 24\%), showing that without structured guidance, LLM often produces subpar fixes – e.g., catch Exception generically or print errors without context, which reviewers frown upon. In contrast, Seeker’s output adhered to best practices (e.g., using specific exceptions, meaningful logging, not altering functionality), earning high praise in automated and manual reviews.
\end{itemize}

\begin{figure*}[!ht]
  \centering
  \includegraphics[width=0.9\linewidth]{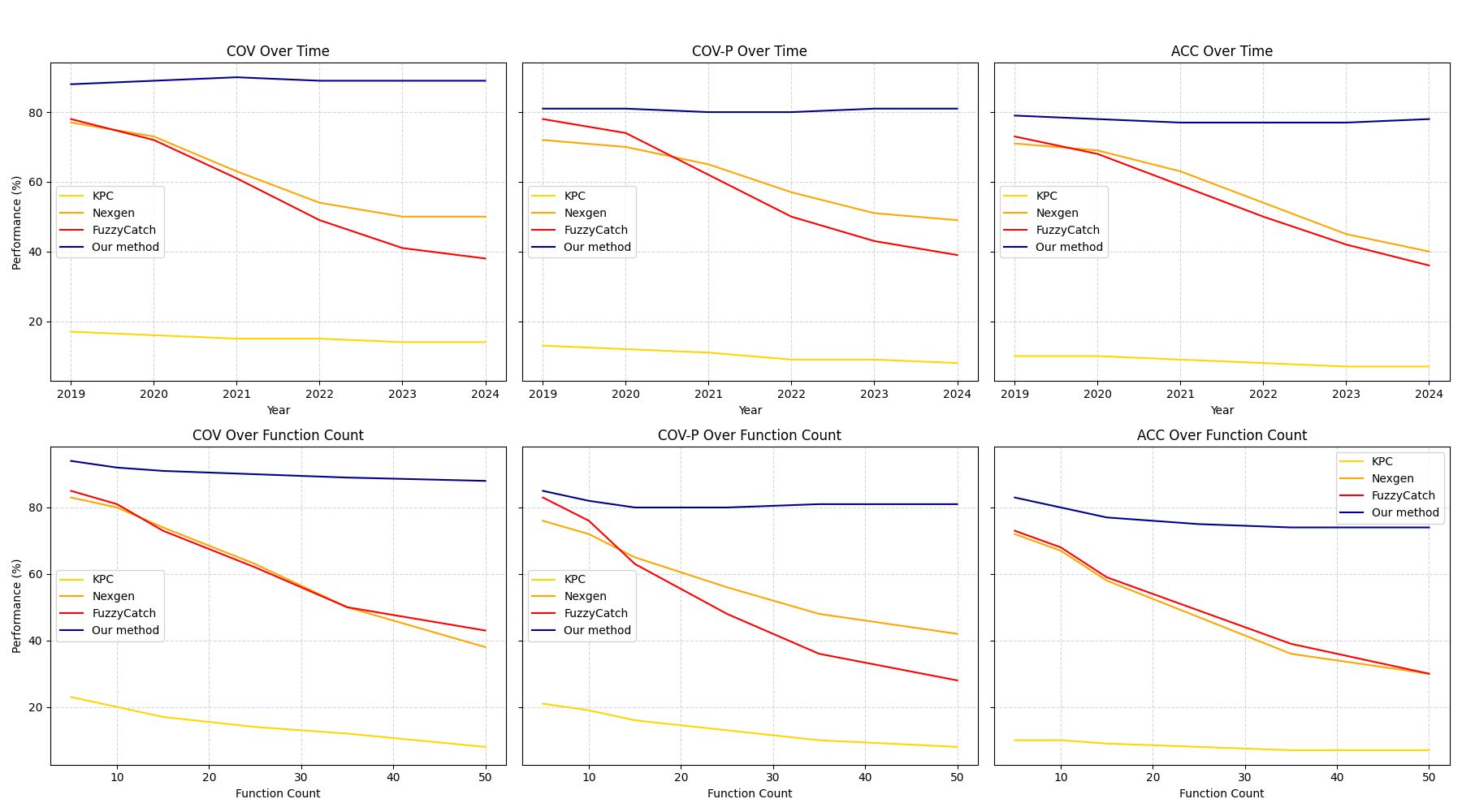}
  \caption{Comparison of Performance Stability Across Baselines and Our Method over Varying Conditions. The top set of curves illustrates the performance metrics over time (2019 to 2024) across different baselines and our method. The bottom set displays performance across increasing function counts.}
  \label{fig8}
  \vskip -.1in
\end{figure*}
\label{sec:RQ1}

Figure \ref{fig8} further illustrates one aspect: stability over time and complexity. We partitioned our test cases by year (to see if newer code is harder due to more modern patterns that older baselines might not know) and by complexity (in terms of function length and number of possible exceptions). We observed that baseline methods often had inconsistent performance – e.g., some had an uptick on older code (maybe because training data had similar code) but dropped on newer code, and all baselines’ performance declined as code complexity increased (multiple functions, nested exceptions, etc.). Seeker, on the other hand, maintained consistently high performance across time periods and complexities. It was less sensitive to such variations, likely because its approach doesn’t rely on seeing exact patterns before – it dynamically analyzes each case. This stability is crucial for real-world applicability, as one wants a tool that works reliably on various projects and doesn’t degrade on larger modules.

\begin{tcolorbox}[size=title]
    \textbf{Answer to RQ1}: \ding{182} The naive approaches (General Prompting, Web-Search RAG) left many issues unhandled (low COV) and often didn’t choose the right solutions (low ACC, low CRS). \ding{183} Tools like FuzzyCatch and Nexgen, while better than naive, still missed many cases or mishandled them. They likely suffer from not generalizing to all scenarios (may have been tuned to specific contexts only). \ding{184} KPC, which is a strong prompt-chaining baseline, didn’t perform well in real-world evaluation (e.g., COV only 14\%). It likely because KPC was tailored to API-specific patterns and the real-world implementation does not match patterns directly \cite{deveval}. It also struggled without certain environment assumptions \cite{metamind}. \ding{185} Seeker’s near-perfect detection and high quality handling show that combining expert knowledge with stepwise IR reasoning yields results that approach human performance.
\end{tcolorbox}

\subsection{RQ2: Effect of Different Agents in Seeker}
\label{sec:RQ2}

We conducted an ablation study to understand how each component of the Seeker framework contributes to its overall performance. We created five ablated versions of Seeker, each with one agent removed:
\begin{itemize}
    \item Scanner: Seeker without the Scanner agent (i.e., not partitioning code into units, treating whole input as one segment).
    \item Detector: Seeker without the Detector (i.e., assume all code needs handling, no selective identification of fragile code).
    \item Predator: Seeker without Predator (i.e., not explicitly enumerating exceptions via Deep-RAG; in this ablation, Detector’s output goes directly to Ranker with full CEE).
    \item Ranker: Seeker without Ranker (i.e., no strategy selection; Predator’s identified exceptions all get handled in a default way).
    \item Handler: Seeker without Handler (i.e., the framework identifies exceptions but doesn’t adapt the code accordingly 
\end{itemize}
The results are presented in Table~\ref{table:ablation}. We see significant drops in performance when any agent is removed, confirming that each plays a critical role:

\begin{table}[ht]
\centering
\caption{Ablation Study on the Effect of Different Agents}
\label{table:ablation}
\resizebox{0.8\columnwidth}{!}{
\begin{tabular}{lcccccc}
\toprule
\textbf{Configuration} & \textbf{ACRS} & \textbf{COV (\%)} & \textbf{COV-P (\%)} & \textbf{ACC (\%)} & \textbf{ES} & \textbf{CRS (\%)} \\
\midrule
\textbf{Seeker (Full)}         & \textbf{0.85} & \textbf{91} & \textbf{81} & \textbf{79} & \textbf{0.64} & \textbf{92} \\
- Scanner       & 0.78 & 85 & 75 & 73 & 0.59 & 86 \\
- Detector      & 0.76 & 63 & 54 & 61 & 0.51 & 84 \\
- Predator      & 0.72 & 61 & 53 & 42 & 0.47 & 81 \\
- Ranker       & 0.63 & 90 & 79 & 75 & 0.49 & 65 \\
- Handler      & 0.50 & 91 & 81 & 79 & 0.34 & 42 \\
\bottomrule
\end{tabular}
}
\end{table}

\begin{itemize}
    \item Without Scanner: Performance declines modestly but noticeably (ACRS 0.78 vs 0.85 full; CRS 86\% vs 92\% full). COV drops from 91\% to 85\%, meaning missing fragile code segments. This suggests that partitioning code helps ensure thorough analysis. Without scanning, the system might be overwhelmed by large context or fail to spot issues in a big function because it’s not focusing line-by-line. The drop in CRS (code quality) could be due to handling too large a chunk at once, possibly leading to less clean outputs.
    \item Without Detector: We see a larger drop in coverage (down to 63\%). By trying to handle everything, the system likely inserted unnecessary try-catches (result in false positives, thus low COV-P 54\%). Effectively, it did not truly cover the right spots well. The quality metrics (ACRS 0.76, CRS 84\%) also suffer because handling code that isn’t actually fragile can introduce clumsy or needless try-catches, lowering code quality. This underscores Detector’s importance in targeting: without it, the framework loses precision and wastes effort on non-issues.
    \item Without Predator: This had a notable effect on ACC (dropping to 42\%) and overall performance (ACRS 0.72). Predator is responsible for identifying specific exception types. Without it, the system likely defaulted to some generic handling, resulting in low accuracy of exception types. Coverage also fell (61\%). Essentially, without Predator’s deep knowledge retrieval, the system doesn’t know what to catch, so it either under-catches or over-catches generically. This highlights Predator (and thus the Deep-RAG + CEE combo it uses) as essential for accuracy and completeness in exception identification.
    \item Without Ranker: Interestingly, COV and ACC remain high (90\% and 75\% respectively, close to full) because Predator still identified the right exceptions and Handler applied them. However, the code quality metrics plummet – ACRS 0.63 (vs 0.85 full) and CRS only 65\% (vs 92\% full). This indicates that while exceptions were caught, the handling strategies were suboptimal without Ranker. Likely, in the absence of Ranker’s strategic selection, the system might have applied default handling for exceptions, which in some cases was not appropriate. For example, it might always log and continue, even when it should have rethrown or returned – things that an intelligent choice would change per context. The Edit Similarity also drops (0.49 vs 0.64), meaning the solutions looked less like human ones (more boilerplate or incorrect style). So Ranker’s role in picking the right handling approach is key to producing high-quality, review-pleasing code.
    \item Without Handler: Here we simulate detecting issues but not actually adapting with the code. As expected, coverage stays 91\% and ACC 79\% (it identified the exceptions). But since no adaptation guidelines pair with the code, the function of the code might not be aligned, leading to an effective code review score of only 42\% and ACRS 0.50. This highlights that the final step of adaptive implementing the changes is necessary.
\end{itemize}
This ablation confirms the synergy of the five agents. Each agent’s output feeds into the next in a way that the whole is greater than sum of parts. For instance, Predator and Ranker together ensure not just correctness but also optimality of the solution – Predator gives options, Ranker chooses the best; you lose a lot if either is absent. The interplay ensures comprehensive coverage (from detection through to implementation) and high-quality outcomes (due to strategic selection and careful code adaptation). As a result of this study, we are confident that our multi-agent design choices were appropriate. It also suggests that if one were to further improve Seeker, each agent offers a point for enhancement (e.g., one could further develop Detector using static analysis, or a better Ranker using learned policy, etc.), but none of them appears redundant.
\begin{tcolorbox}[size=title]
    \textbf{Answer to RQ2}: Each agent contributes uniquely to Seeker’s performance: \ding{182} Scanner ensures no fragile piece is overlooked and helps manage complexity, improving overall quality. \ding{183} Detector provides focus, ensuring precision so that only actual issues are handled (improving COV-P and avoiding quality degradation from false positives). \ding{184} Predator supplies the domain knowledge to catch the correct exceptions, crucial for ACC and coverage. \ding{185} Ranker fine-tunes the solution to be appropriate and elegant, greatly affecting code quality and acceptance. \ding{186} Handler actually adapts the fix in code, without which all else would be abstract theoretical.
\end{tcolorbox}

\subsection{RQ3: Effect of Underlying Language Model}
\label{sec:RQ3}
Seeker’s design is modular with respect to the underlying LLM – the agents can in theory use any language model as their reasoning/generation engine. In this experiment, we test Seeker with different LLMs to see how they influence performance. We consider both open-source models and closed-source models, Table \ref{table:leaderboard} shows the results:
\begin{table}[ht]
\centering
\caption{Performance of Different Models on Exception Handling Code Generation}
\label{table:leaderboard}
\resizebox{0.8\columnwidth}{!}{
\begin{tabular}{lcccccc}
\toprule
\textbf{Model} & \textbf{ACRS} & \textbf{COV (\%)} & \textbf{COV-P (\%)} & \textbf{ACC (\%)} & \textbf{ES} & \textbf{CRS (\%)} \\
\midrule
\multicolumn{7}{c}{\textbf{Open-Source Models}} \\
\midrule
Code Llama-34B      & 0.31  & 37  & 35 & 32  & 0.25 & 34  \\
WizardCoder-34B     & 0.37  & 35  & 31  & 29  & 0.28 & 35  \\
Vicuna-13B           & 0.23  & 15  & 9  & 11  & 0.19 & 26 \\
\midrule
\multicolumn{7}{c}{\textbf{Closed-Source Models}} \\
\midrule
Claude-2           & 0.42  & 64  & 59  & 54  & 0.40 & 54  \\
GPT-3-davinci        & 0.56  & 78  & 68  & 60  & 0.48 & 58  \\
GPT-3.5-turbo       & 0.63  & 79  & 72  & 66  & 0.52 & 71  \\
GPT-4-turbo         & 0.84  & \textbf{91}  & \textbf{83}  & 77  & 0.63 & 89  \\
GPT-4o        & \textbf{0.85} & \textbf{91} & 81 & \textbf{79} & \textbf{0.64} & \textbf{92} \\
\bottomrule
\end{tabular}
}
\end{table}

\ding{182} The open-source models underperform significantly compared to close-source models. Code Llama and WizardCoder managed ACRS around 0.3–0.37 and low coverage (35–37\%). These models missed many exceptions (COV < 40\%) and had low accuracy in type selection. Likely reasons are: [1] although open-source models are fine-tuned with code, due to the exception defects in training data, their exception understanding and generation capabilities are weak. [2] due to the lack of general knowledge, they might not follow the multi-step instructions of our agents as reliably (we observed sometimes they did not follow the plan correctly, or produced incoherent outputs for Handler).

\ding{183} The closed-source models show a clear progression with capability. Claude-2 (outperform monolithic GPT-4o level) got ACRS 0.42, better coverage (64\%) than open models but still far from GPT-4. GPT-3’s old davinci model did okay (ACRS 0.56, COV 78\%), showing that even older OpenAI models had potential competency. GPT-3.5-turbo improved further (ACRS 0.63, COV 79\%), though not nearly as precise or high-quality as GPT-4 (CRS 71\% vs 89\%). It also suggests that GPT-4-level model suffices to realize the full potential of our approach.

\ding{184} Models with larger pre-training on general knowledge (GPT series) clearly knew the common exception patterns better. For example, GPT-3.5 and above rarely missed adding a catch for obvious exceptions, whereas open models sometimes didn’t put any try-catch for cases. We also found smaller models often gave irrelevant or hallucinated answers (Vicuna sometimes listed exceptions that were not applicable at all). Larger models stuck to relevant ones, showing better comprehension. 

\ding{185} The retrieval part (Deep-RAG + CEE) provides the model info like “Exception X can occur, recommended to do Y”. We saw GPT-4 uses that info smartly in its output, but models like Code Llama might simply regurgitate part of it without applying correctly, or ignore some of it. The better the model, the better it utilizes retrieved knowledge.

We note that even the best open-source model (WizardCoder-34B) had performance similar to GPT-3 from years ago. This suggests that for now, closed models like GPT-4 still have a considerable edge for complex tasks like this. On the positive side, Seeker with GPT-3.5-turbo already surpasses most baselines and could be considered usable. 

\begin{tcolorbox}[size=title]
    \textbf{Answer to RQ3}: The underlying model’s strength significantly influences Seeker’s results due to training data, post-training method, model architecture \& size, and integration of RAG. But we elevate all models to do better at exception handling than they would with naive prompting. We recommend using the best available model for critical applications, but even in resource-limited settings, the structured approach yields improvements. And as models improve, we believe Seeker will harness those effectively.
\end{tcolorbox}

\subsection{RQ4: Impact of Domain-Specific Knowledge Integration}
\label{sec:RQ4}

One of the core hypotheses in Seeker is that integrating an external knowledge base (CEE) of exception handling technical practices significantly boosts performance. To validate this, we ran Seeker in two modes on the primary dataset: with CEE (the full system) and without CEE. Without CEE means the Predator and Ranker agents did not have access to the curated knowledge – Predator would just rely on the base model’s intuition to analyze exceptions, and Ranker wouldn’t have standardized strategies to choose from. The results are presented in Table \ref{table:cee}.

\begin{table}[h]
\centering
\caption{Impact of Integrating Common Exception Enumeration (CEE)}
\label{table:cee}
\resizebox{0.8\columnwidth}{!}{
\begin{tabular}{lcccccc}
\toprule
\textbf{Configuration} & \textbf{ACRS} & \textbf{COV (\%)} & \textbf{COV-P (\%)} & \textbf{ACC (\%)} & \textbf{ES} & \textbf{CRS (\%)} \\
\midrule
\textbf{Seeker}     & \textbf{0.85} & \textbf{91} & \textbf{81} & \textbf{79} & \textbf{0.64} & \textbf{92} \\
- CEE       & 0.38 & 48 & 41 & 32 & 0.29 & 46 \\
\bottomrule
\end{tabular}
}
\end{table}
The difference is striking. With CEE, Seeker achieves top-tier performance. Without CEE, its performance plummets across all metrics:

\begin{itemize}
    \item ACRS drops from 0.85 to 0.38 – meaning overall code quality and correctness are far lower without the knowledge base.
    \item Coverage falls to 48\% (from 91\%) - meaning the system without CEE is detecting less than half of the issues it should, indicating that without the knowledge base to guide what to look for, many fragile spots go unnoticed.
    \item COV-P similarly at 41\% (vs 81\%) – showing that even the segments it does handle are handled incorrectly. The Predator likely mispredicts exceptions or highlights wrong operations when blind.
    \item ACC plummets to 32\% (from 79\%) – showing that without CEE, the exceptions caught were likely wrong or too generic. We observed that LLM often resorted to catching broad exceptions or missed specific ones entirely. This underscores how much CEE contributes to knowing which exception types are relevant.
    \item ES at 0.29 (vs 0.64) – the changes are far less similar to human fixes because they are either simplistic or misguided.
    \item CRS at 46\% (vs 92\%) - showing that only about half of the code reviews pass when CEE is removed, i.e., many mistakes or omissions remain, which a reviewer would flag.
\end{itemize}
These numbers confirm that domain-specific knowledge (CEE) yields substantial improvements across the board. To give an intuitive example from our tests: consider a code that interacts with a database. Without CEE, the model does not recall exceptions like \texttt{SQLException} or just catch \texttt{Exception} and print. With CEE, Predator explicitly brings up \texttt{SQLException} and its subclasses, and suggests strategies like "catch \texttt{SQLTimeoutException} separately if it’s a timeout, etc." As a result, with CEE the final code had fine-grained catches and proper logging; without it, the code often either missed catching or did a generic catch that got low marks. We also found that even without CEE, Seeker (with chain-of-thought) slightly outperforms the naive LLM prompting – for instance, comparing no-CEE Seeker (ACRS 0.38) to the General Prompting baseline (0.21) despite of CEE provides the crucial content for those steps.

\begin{tcolorbox}[size=title]
    \textbf{Answer to RQ4}: CEE is a game-changer for Seeker’s effectiveness. It supplies the model with a rich context that it doesn’t reliably hold in its parameters. Many of the details in CEE might be things an LLM could theoretically learn from training (like common exceptions for certain operations), but our results indicate that LLMs benefit hugely from having this information explicitly provided. \ding{182} It reduces omissions: LLMs might not recall a rare exception type at generation time without prompting. CEE ensures it’s listed. \ding{183} It standardizes quality: LLMs might know an exception, but not how to handle it well. CEE gives it a high-quality handling principle, so it doesn’t do something ad-hoc or wrong. \ding{184} It improves confidence: The model might be unsure if it should catch something or not; seeing it in CEE confirms that it’s expected to catch it, so it does.
\end{tcolorbox}

\subsection{RQ5: Additional Analysis}
\label{sec:RQ5}
Beyond our primary evaluation on exception handling tasks, we wanted to assess Seeker’s applicability in more complex, real-world scenarios and standard code generation benchmarks. We present two analyses: applying Seeker to [1] a software bug fixing benchmark (SWE-bench) and [2] a pragmatic code generation benchmark (CoderEval). These tests demonstrate Seeker’s generality and the incremental improvements it can provide when robust exception handling is crucial.

\subsubsection{Application to SWE-bench}
SWE-bench \cite{swebench} is an evaluation framework that consists of 2,294 real GitHub issues along with the corresponding patches (fixes) in 12 popular Python repositories. The tasks require a model to modify a given codebase to resolve a described issue, which often involves coordinating changes across multiple files and understanding long contexts. This benchmark is challenging on not just writing a function, but understanding an issue, running tests, etc. The results were measured in terms of:
\begin{itemize}
    \item Resolve Rate: the percentage of issues for which the model’s changes completely solved the problem (based on tests and criteria from the benchmark).
    \item Apply Rate: the percentage of the model’s patches that could be applied to the codebase without causing errors (based on whether it produce a valid patch that integrates without breaking things, even if it may not fully solve the issue).
\end{itemize}

We used SweAgent \cite{yang2024sweagentagentcomputerinterfacesenable}, which is an agent-based system for automated issue resolution coupled with GPT-4o, as a baseline to solve these issues. We applied our Seeker-Python to attempt the same issues, the results are presented in Table \ref{table:swebench_results}.

\begin{table}[ht]
\centering
\caption{Performance on SWE-bench Solving Real Development Issues}
\label{table:swebench_results}
\resizebox{0.6\columnwidth}{!}{
\begin{tabular}{lcc}
\toprule
\textbf{Method} & \textbf{Resolve Rate (\%)} & \textbf{Apply Rate (\%)} \\
\midrule
SweAgent + GPT-4o & 19.10 & 43.56 \\
\textbf{Seeker} + GPT-4o & \textbf{27.98} & \textbf{62.11} \\
\bottomrule
\end{tabular}
}
\end{table}

\begin{itemize}
    \item Resolve Rate: SweAgent solved about 19.10\% of issues, whereas Seeker solved 27.98\%. This is a relative improvement of 46\% in success rate. Considering these are real issues, an 8.9 percentage increase is meaningful.
    \item Apply Rate: SweAgent had 43.56\% of its patches apply successfully, while Seeker had 62.11\%. So Seeker’s patches were not only more often correct, but also more often syntactically/semantically valid (didn’t introduce conflicts or errors). The ~18.5 point jump here suggests Seeker’s structured approach yields changes that integrate more cleanly with the existing codebase. This could be because exception handling improvements are often additive and less likely to conflict with logic, whereas the baseline might make riskier changes.
\end{itemize}
These results demonstrate Seeker’s benefit in a practical scenario: improving existing code. Many issues in SWE-bench revolve around things like unhandled exceptions causing program crashes or user-facing errors – exactly what Seeker is designed to fix that the general agent missed. Also, because Seeker focuses on not breaking functionality, its patches were more likely to be acceptable.
\subsubsection{Application to CoderEval}
CoderEval \cite{codereval} is a benchmark for pragmatic code generation, meaning tasks that involve writing code which often interacts with other functions or requires handling of external resources. It moves beyond single isolated functions. The key metric in CoderEval is Pass@1 – the percentage of tasks where the model’s first attempt is a correct solution (runs and produces expected output). Many tasks here involve writing code to specification, which can include internal error handling or working with tricky inputs. The results are presented in Table \ref{table:codereval_results}.

\begin{table}[h]
\centering
\caption{Performance on CoderEval Java Code Generation Tasks}
\label{table:codereval_results}
\begin{tabular}{lc}
\toprule
\textbf{Method} & \textbf{Pass@1 (\%)} \\
\midrule
Codex & 27.83 \\
\textbf{Seeker} + Codex & \textbf{38.16} \\
\bottomrule
\end{tabular}
\end{table}

The integration of Seeker improved Codex’s performance from 27.83\% to 38.16\% Pass@1. That’s a substantial improvement on an already fairly challenging benchmark. We attribute this to the fact that:

\ding{182} Some tasks require robust handling of input or multiple scenarios to pass all test cases. Codex alone fails on edge cases \cite{deveval} (like not handling when an input is null or out-of-range). Seeker’s guidance likely caused Codex to add the necessary checks or try-catches, thus passing more tests.

\ding{183} The structured approach could have reduced logic errors. By thinking in terms of scanning, detecting issues, etc., even if implicitly through our prompts, Codex might have produced more logically sound code.

\ding{184} Also, since Codex is a bit older, providing it with the extra knowledge via CEE has helped cover cases it wasn’t trained heavily on \cite{li2024evocodebench}.

These results underscore that Seeker is not limited to contrived scenarios; it has real impact on broader coding tasks. By plugging into existing code generation pipelines, it can enhance reliability and correctness across diverse problems. Also, inspired by SocialEval \cite{zhou2025socialeval} and DoT \cite{dot}, we found that Seeker framework has more room for development in general LLM reasoning. Through pre-deduction in tree inference, LLM is expected to enter the problem-solving ideas more efficiently and optimize its reasoning actions through interaction with the external environment. A potential application is shown in Figure \ref{fig8.5}. In the future, we will continue to explore research in this direction.

\begin{tcolorbox}[size=title]
    \textbf{Answer to RQ5}: When exception handling quality matters (which is often in real projects), Seeker’s techniques yield measurable improvements. Whether it’s fixing real bugs or writing new code that’s resilient, the multi-agent IR approach proves beneficial. 
\end{tcolorbox}

\section{Discussion}
\label{sec:discussion}

\begin{figure}[t]
\centering
\includegraphics[width=0.7\linewidth]{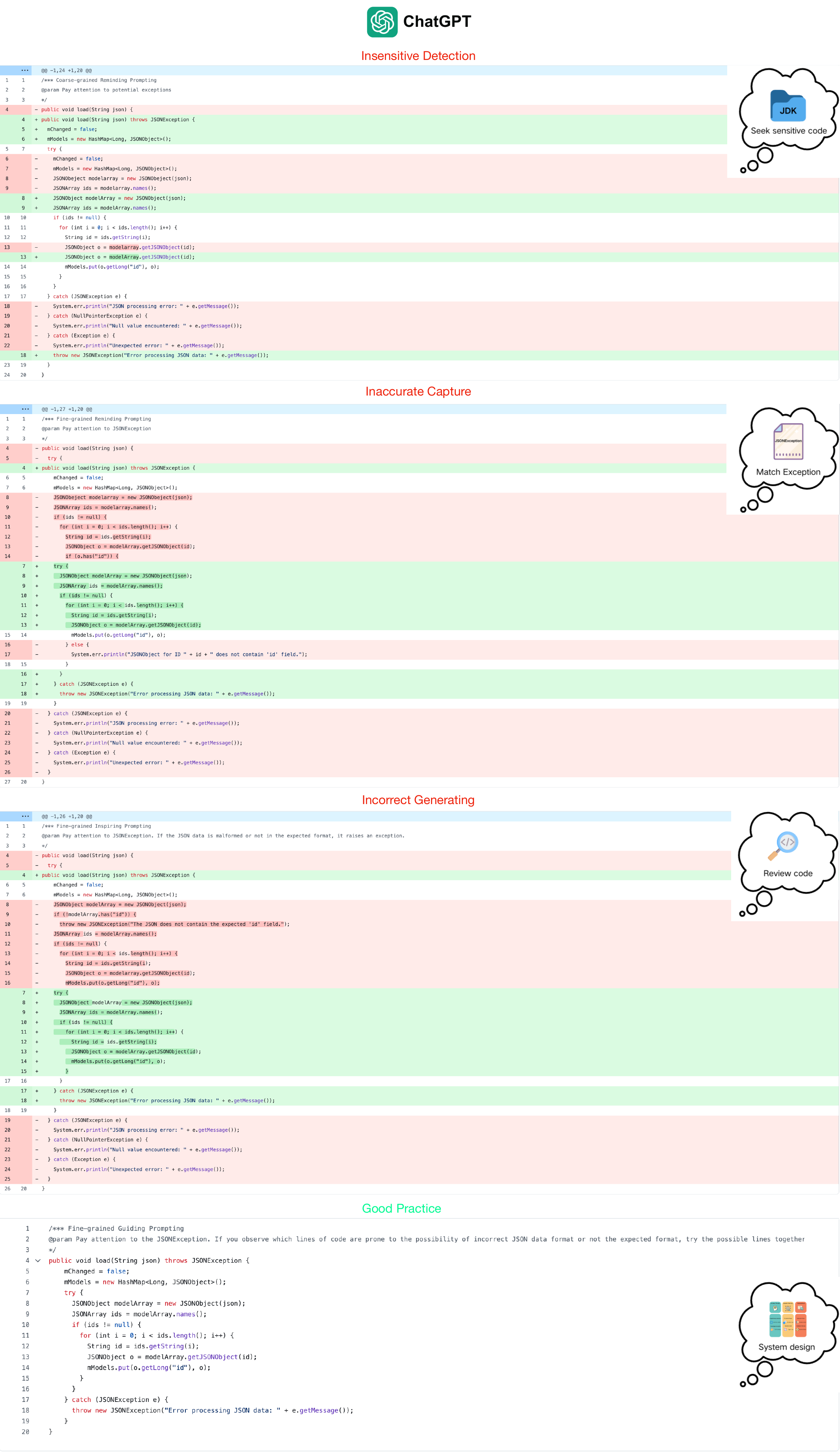}
\caption{A schematic illustration of the preliminary phenomenon, showing how incremental, targeted guidance enhances LLM-based exception handling. The depicted code segments and annotations highlight which specific information supports more accurate detection and handling of fragile code scenarios.}
\label{fig2.1}
\end{figure}

\begin{figure}[t]
\centering
\includegraphics[width=0.8\linewidth]{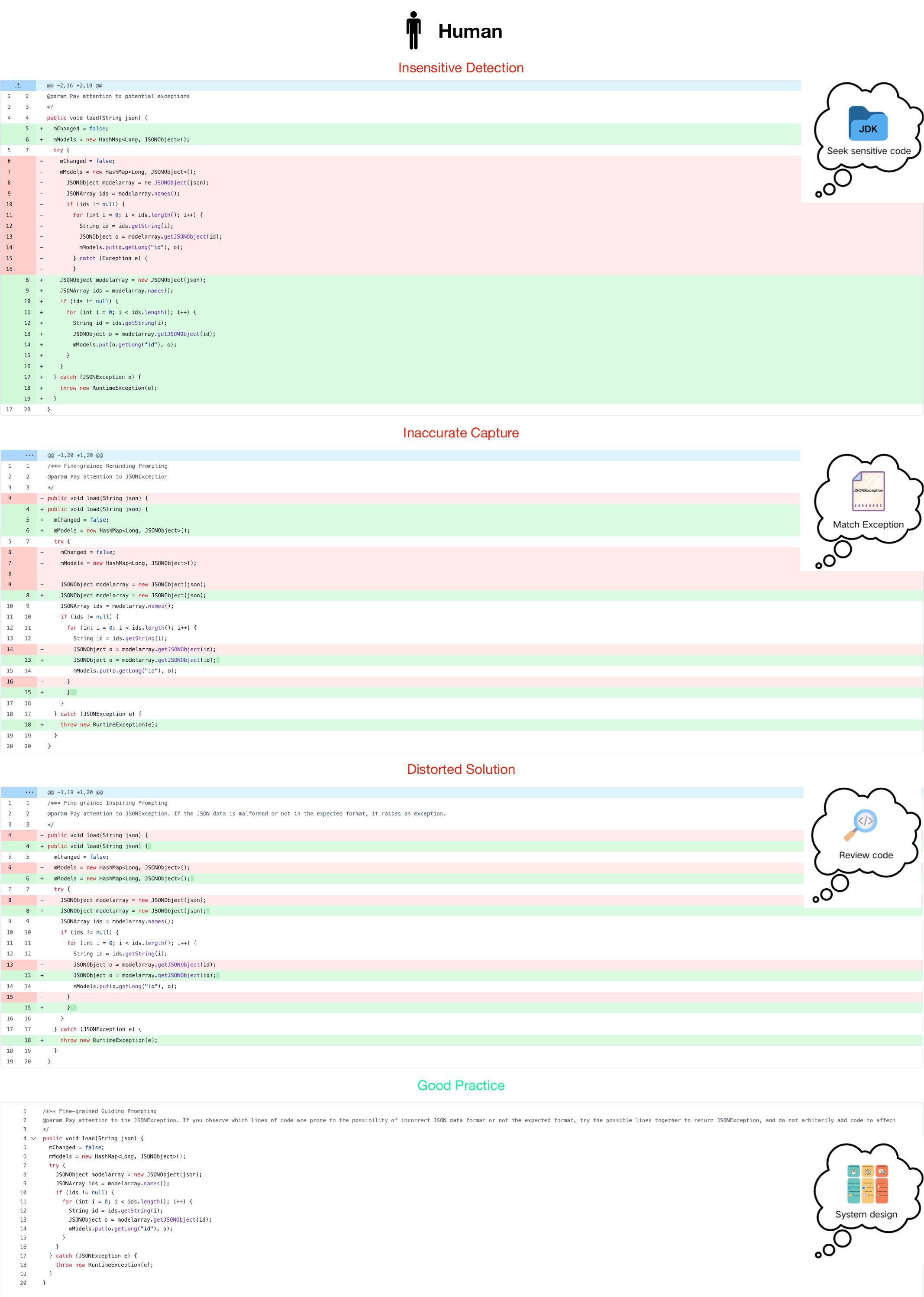}
\caption{A schematic illustration of the preliminary phenomenon, demonstrating that incremental, targeted guidance similarly benefits both LLMs and human developers in exception handling. The highlighted case study underscores which information elements help bridge the gap between current human practice and reliable, automated handling strategies.}
\label{fig2.2}
\end{figure}
In designing and evaluating Seeker, we made certain choices about scope and technique that warrant further discussion. We elaborate on two key points – why we focused on Java as the initial target language, and why we emphasize try-catch blocks as the means of exception handling – and then consider the broader implications of our results for software engineering and AI-assisted programming.
\subsection{Why Java?}
% \paragraph{}
We chose to first implement and test Seeker for the Java programming language due to both practical need and technical challenge. Java’s exception handling model is one of the most rigorous among mainstream languages, featuring a mix of checked and unchecked exceptions and a deeply nested inheritance hierarchy (over 433 exception classes). This complexity means that Java developers often struggle with exception management, and projects are prone to bugs arising from poor handling. Indeed, studies have shown that Java projects accumulate exception-related bugs (like misused or missing catches) over time \cite{Java}. By targeting Java, we address a space where the need for improved exception handling is urgent – many large enterprise systems rely on it and could benefit from such automation.

From a technical standpoint, Java presents the most challenging test case for exception handling approach. \ding{182} The large exception hierarchy (433+ nodes, 5+ levels) stresses Deep-RAG algorithm’s ability to efficiently retrieve relevant exceptions. If Deep-RAG can handle Java’s tree, it can likely handle simpler ones (e.g., Python’s). \ding{183} Java’s language rules around exceptions (checked exceptions must be declared or caught) mean that missing a catch can break compilation. Thus, a system needs to be thorough and correct – an ideal proving ground for Seeker’s thoroughness. \ding{184} Java code often uses exceptions as part of normal control flow (for example, iterators throw \texttt{NoSuchElementException} to signal end of iteration). Handling such situations properly is tricky and requires knowledge of best practices.

And indeed, our design is language-agnostic at the architecture level – the agents remain the same; one only needs to swap out the parsing rules and the knowledge base for a new language. For example, a Python version would need a Python CEE (which is much smaller since Python has fewer built-in exceptions) and minor adjustments in how try/except is generated for the Handler agent. We believe our results would translate to languages like C\#, JavaScript etc. with at most a moderate effort of retuning, thanks to the modular design of Seeker. As mentioned, we have already tested Seeker-Python and will release soon.
\subsection{Why Try-Catch?}
% \paragraph{}
Exception handling can be done via different mechanisms in various languages. In Java and many others, there are typically three ways: \ding{182} Declaring the exception in the method signature (using \texttt{throws} in Java) so callers know it might happen. \ding{183} Throwing the exception up (using \texttt{throw}) intentionally after wrapping it. \ding{184} Capturing it via a \texttt{try-catch} block and dealing with it on the spot.

We centered Seeker’s fixes around the try-catch approach, especially in Java, for several reasons:

\begin{itemize}
    \item Runtime Robustness: Catching exceptions where they occur (or at an appropriate boundary) ensures that the program can gracefully handle the error and continue or terminate safely. Declaring with \texttt{throws} simply punts the problem to a higher layer; if that layer doesn’t handle it, the program still crashes. Our goal was to proactively embed robustness, and try-catch does that by actually intercepting the exception flow and dealing with it.
    \item Maintainability: Encapsulating error handling logic near the source of the error can make code more self-contained and easier to reason about. If you look at a method and see how it handles its potential errors, you don’t need to trace as much into callers. Using throws defers that, which sometimes is fine (for library code), but often just burdens the next layer. Empirical work \cite{dont-coarse} noted that liberally using \texttt{throws} doesn’t reflect true runtime conditions – eventually something up the chain must catch it, and often by then it’s unclear what to do. Our approach encourages handling sooner rather than later when appropriate.
    \item Technical Practices Alignment: Industry practice in Java leans towards using checked exceptions for recoverable conditions that you should catch, and using runtime exceptions for programming errors that often crash. We integrated this idea: e.g., Seeker doesn’t try to catch things like \texttt{OutOfMemoryError} or \texttt{AssertionError} – those are better left to propagate (or crash) since you can’t meaningfully handle them. But it will catch things like I/O errors. This aligns with how robust Java programs are written: catch what you can handle, declare or propagate what you cannot. We found try-catch to be the most practical and common approach for managing errors, especially at application boundaries.
    \item Generality of Prompting: Our intermediate representation prompt strategy was naturally suited to instruct the model to add try-catch blocks. For instance, “handle possible errors here” directly implies a try-catch addition. Guiding a model to use \texttt{throws} declarations is less straightforward, and in many cases do not solve the problem (just move it). Meanwhile, try-catch has a direct effect on the code flow, which the model can simulate and reason about in our chain of prompts.
\end{itemize}
That said, try-catch is not a panacea. Overusing it can lead to swallowed exceptions or messy logic, but when in doubt, adding a catch with at least a log and safe recovery is better than leaving a crash. We ensured through our knowledge base that we promote responsible use of try-catch.

It’s worth noting that languages like C++ have alternatives (error codes, optional types, etc.), and some newer languages or frameworks prefer monadic error handling (like Go’s \texttt{error} returns). Our approach could in theory be adapted to those styles by changing what the “Handler” does (e.g., instead of a try-catch, returning an error code or wrapping in \texttt{Result} type in Rust, etc.). But in languages that support exceptions, try-catch remains the idiomatic way to handle them. For our target domain (improving existing code), inserting try-catch blocks is the least intrusive fix – it doesn’t change the method’s signature or the control flow for normal operation, only adds an alternate path for error cases. This fits well with our principle of preserving functional correctness.

Looking ahead, one could extend Seeker to suggest other patterns (like using try-with-resources for certain cases, which is essentially a variant of try-catch for resource management). Our knowledge base in fact already includes hints for that (like for I/O streams, it might prefer try-with-resources to ensure closure). So, try-catch is not just about catching, but also about ensuring cleanup and safety, which our system considers.

\subsection{Threats to validity}
We now discuss potential threats to the validity of our results and how we mitigated them:

\begin{itemize}
    \item \textbf{Base Model Dependency:} One threat is that Seeker’s performance depends on the underlying LLM’s understanding of code. Our evaluation showed that using a weaker base model resulted in lower performance (Section RQ3). This means our claims about improvements are contingent on having a sufficiently capable base model. If the base model doesn’t understand code or instructions well, Seeker would underperform. We mitigated this by testing a range of models and reporting those results. For practical usage, one should use Seeker with a strong base model (ideally GPT-4 level or above) to achieve good results.
    \item \textbf{Closed-Source Model Bias:} We used closed-source models (GPT-4, etc.) extensively. These models, while powerful, may have hidden biases or limitations (e.g., they might have seen some of our test code if it was public, or they might have specific failure modes on certain patterns). We attempted to mitigate data leakage issues: our dataset was drawn from 2019–2024 code and issues that are not part of GPT-4’s training (which cutoff around 2021-2022 for most data). There’s a possibility that GPT-4 had seen some of the StackOverflow or typical solutions for exceptions, giving it an edge. However, the dramatic difference between using that knowledge via our framework vs. the baseline prompts suggests it’s more about how we guided it than it having memorized answers.
    \item \textbf{Private Code Patterns:} Our evaluation datasets are from open-source projects and certain well-known domains. Proprietary or domain-specific code might have different exception handling needs or patterns (for instance, a financial system might have custom exception classes for business logic). Our CEE does not cover those out-of-the-box, which would be a threat to generalizability. To mitigate, our knowledge base structure allows adding new exceptions easily; a user in a proprietary environment could extend CEE with their custom exceptions. We have not yet evaluated Seeker on proprietary code, which is future work. But the architecture is flexible to adaptation with additional tuning or knowledge injection.
    \item \textbf{ACRS Metric and Code Review Emulation:} We introduced the ACRS (Average Code Review Score) metric that involves some level of subjective judgment (even though we used a consistent rubric and partly automated it). There is a threat that our scoring (especially via an LLM “code reviewer”) might be biased or not fully representative of human preferences. We addressed this by cross-checking a subset of outputs with human experts – the correlation was high between our automated review scores and human opinions. We also anonymized and removed any project-specific code review rules in calculating ACRS to avoid bias. Still, code review scoring can be subjective, so we mainly rely on the more objective metrics (coverage, accuracy) for claims, using CRS/ACRS as supportive evidence of code quality improvements.
\end{itemize}
In summary, while our controlled experiments show clear benefits, deploying Seeker in varied real-world settings may encounter conditions we haven’t covered (different coding styles, frameworks, or model constraints). Our mitigation strategies – thorough multi-model testing, flexible knowledge base design, and verification of metrics with human judgment – provide reasonable confidence in our results. Nonetheless, ongoing evaluation in industry settings (with feedback from developers) will be important to fully validate Seeker’s effectiveness and adapt it to any unforeseen challenges.

\begin{figure*}[h]
  \centering
  \includegraphics[width=0.9\linewidth]{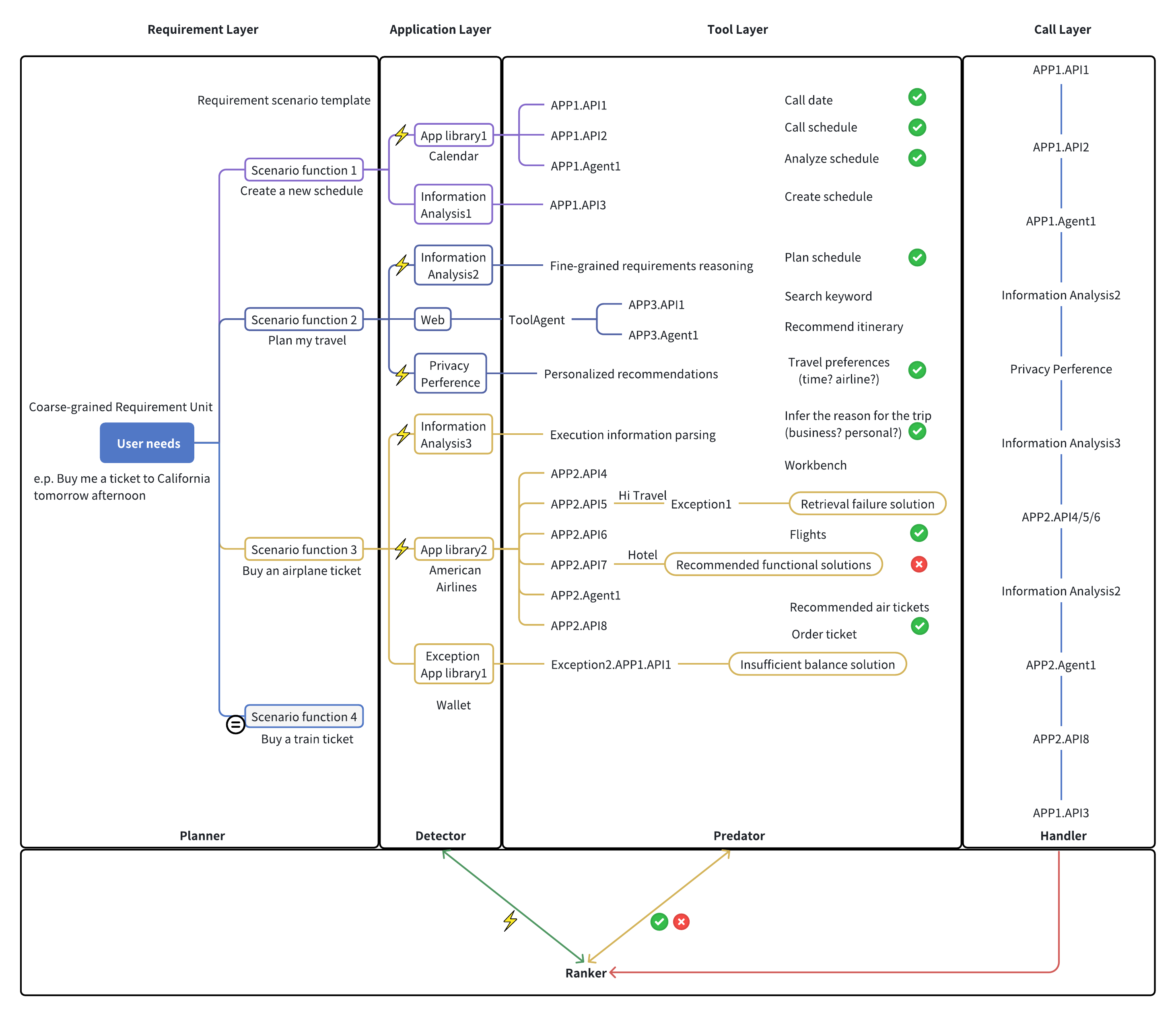}
  \caption{A schematic depiction of integrating the Seeker multi-agent framework into APP requirement engineering workflows. By bridging layered requirements, application functionalities, tool integrations, and call-level operations, Seeker generalizes beyond isolated exception handling to more complex inheritance relationships. This approach improves interpretability, scalability, and reasoning capabilities, demonstrating the framework’s adaptability and high performance across diverse, real-world engineering scenarios.}
  \label{fig8.5}
  \vskip -.1in
\end{figure*}

\section{Related Work}
\label{rw222}

Exception handling has long been recognized as vital for software robustness. Nevertheless, developers often struggle with it, and traditional code generation techniques have only partially addressed the challenge. Seeker aims to set a new direction for safer AI-driven code generation, where robustness and reliability are treated as first-class goals alongside functional correctness. We situate our work in three relevant areas: [1] automated exception-handling tools, [2] multi-agent collaboration frameworks, and [3] robust code generation and repair techniques.

\subsection{Automated Exception Handling Tools}

A number of approaches have attempted to automatically suggest or insert exception handling code. Early work includes learning from repositories to recommend exception-handling code snippets for given contexts. For example, FuzzyCatch \cite{baseline3} is a tool that uses heuristics and fuzzy logic to recommend how to handle exceptions in a given code snippet. While it provides suggestions, its scope is limited to patterns seen during training and it may misfire in novel scenarios. Nexgen \cite{baseline1}, a neural pre-training approach, trains on code to learn how exceptions were handled, but it struggles to generalize beyond its training distribution and does not incorporate external knowledge or reasoning. KPC (Knowledge-driven Prompt Chaining) \cite{kpc} is a recent state-of-the-art method for enhancing LLMs in exception handling tasks. KPC uses a series of tailored prompts (especially API-specific prompts) to coax the model into better handling, but it falters with complex codebases that involve multiple interacting exceptions or those outside the APIs it knows. Including traditional static analysis method, common limitations of these tools are limited generalizability (only specific patterns or languages), reliance on training data that may not cover long-tail exceptions (leading to biases) \cite{gptj}, and lack of proactive error mitigation (most focus on reacting to known issues rather than preventing misuse) \cite{final}. In contrast, Seeker addresses these gaps by using a language-agnostic multi-agent framework (enabling cross-language application) and an external knowledge base (CEE) to handle even rare exceptions. Moreover, Seeker emphasizes LLM-guided reasoning instead of solely data-driven pattern matching, which allows it to adapt to novel exception types and contexts while preserving program semantics.

\subsection{Multi-Agent Collaboration}
Orchestrating multiple specialized agents (or AI models) has shown promise in decomposing complex tasks and improving performance. Approaches like VisualGPT \cite{visual} and HuggingGPT \cite{hug} use an LLM as a controller to manage other AI models for multimodal tasks. CAMEL \cite{camel} demonstrates inter-LLM collaboration by simulating roles for two chatbots to cooperate on tasks. In software engineering, CodeAgent \cite{zhang} integrates various tools (like static analyzers, test runners) with LLMs to solve repository-level coding challenges. These works indicate that specialized agent roles and a central coordinator can yield better results than a monolithic model. Our Seeker framework is inspired by this paradigm: we deploy five specialized agents that communicate through an intermediate representation of the code’s exception behavior. Unlike existing multi-agent setups that mostly focus on breaking down functional requirements or combining modalities, Seeker is unique in targeting a non-functional requirement: it focuses on robustness and safety aspects (exception handling) that cut across normal code generation tasks. Additionally, prior multi-agent systems in coding often prioritize generating functionality from requirements and neglect error handling (which is considered ancillary). For example, Self-collaboration Code Generation \cite{dong} coordinates ChatGPT instances to generate code from specs, but it doesn’t address exception safety explicitly. Seeker fills this critical gap by leveraging the strengths of multi-agent collaboration – such as modularity, parallelism, and specialization – and combines it with an external knowledge repository (CEE) to guide the agents. This results in a system that not only generates correct functionality but also proactively embeds robustness checks. To our knowledge, Seeker is the first to integrate a multi-agent LLM framework with a domain-specific knowledge base for the explicit purpose of improving code robustness.

\subsection{Robust Software Development and Repair}
Our work is also related to automated program repair and software robustness enhancement. Traditional program repair tools (e.g., Devign \cite{v2}, and Magis \cite{a2}) often operate by detecting vulnerabilities or bugs and then fixing them via learned transformations. These approaches are typically reactive, addressing issues after they manifest, and they risk altering core functionality or introducing new issues \cite{a3, a1}. Exception handling, on the other hand, is a form of proactive robustness: by adding proper try-catch blocks and checks, we aim to prevent crashes or data corruption before they occur. Prior research has shown that proactively handling exceptions can prevent resource leaks and undefined behaviors \cite{dont-coarse,exception2}, but this area is underexplored compared to post-failure bug fixing. Our approach can be seen as bridging program repair with code generation: Seeker’s Handler agent essentially “repairs” the code by inserting error handling during code generation rather than after the fact. Moreover, classic static analyses identified patterns of exception misuse and proposed ways to detect them, but they did not have a good solution to generate fixes. Seeker builds on the understanding from those works (for example, the dangers of generic or empty catches) and automatically generates best-practice fixes in context, guided by CEE. Notably, Seeker emphasizes maintaining functional correctness while adding exception safety – a principle akin to “do no harm” in program repair. By focusing on try-catch based handling, we also avoid strategies that simply propagate exceptions without resolution.

\section{Conclusion and Future Work}
\label{sec:conclusion}
This work explored the impact of structured prompt specifications and multi-agent collaboration on the robustness of LLM-generated code, specifically targeting exception handling – a critical aspect of software reliability. We proposed Seeker, an intermediate-representation agent framework that significantly improves the exception safety of generated code by orchestrating LLMs through a series of specialized tasks with the support of external knowledge base. 

Through extensive experiments, we first confirmed that guiding LLMs with fine-grained, structured prompts (inspired by expert reasoning) has a clear mitigating effect on poor exception handling practices. Building on this insight, we introduced Seeker: a five-agent system (Scanner, Detector, Predator, Ranker, Handler) that injects expert knowledge and step-by-step reasoning into code generation. A central contribution is the development of CEE documents and Deep-RAG algorithm, which together equip the LLM with an on-demand understanding of language-specific exception hierarchies and handling strategies. With these tools, Seeker transforms the code generation process into one that not only produces functional code but inherently handles errors gracefully. 

Our evaluation demonstrated that a Seeker-augmented GPT-4o achieves SOTA performance in exception handling tasks, surpassing prior methods by large margins. Specifically, Seeker improved exception handling precision, coverage, and code quality, also proved effective across various scenarios. We hope that our findings and the Seeker framework provide new insights into engineering LLM prompts and agents for software quality, suggesting a path forward where LLM-based code generation is not used in isolation, but in tandem with curated domain knowledge and systematic reasoning procedures. Such an approach can greatly enhance the trustworthiness of AI-generated code, making it more feasible for integration into real-world development workflows. 

In future research, we aim to explore additional domains (security, performance) where intermediate representation agents like Seeker can guide LLMs to meet specific non-functional requirements. We also plan to collaborate with the community to expand the CEE knowledge base and evaluate Seeker in industry settings. By continuing to blend human expertise with AI capabilities, we envision tools that not only generate code but also inherently enforce best practices, leading to a new generation of AI-assisted programming that is robust, reliable, and aligned with the standards of professional software engineering.

\begin{acks}

\end{acks}

\bibliographystyle{ACM-Reference-Format}
\bibliography{sample-base}

\appendix

\end{document}